\documentclass[AMA,STIX1COL]{WileyNJD-v2}

\usepackage{amsmath}
\usepackage{mathtools}
\DeclarePairedDelimiter{\ceil}{\lceil}{\rceil}
\usepackage{pifont}
\newcommand{\xmark}{\ding{55}}
\usepackage{algorithm}
\usepackage{algpseudocode}
\usepackage{algorithmicx}
\usepackage{upgreek}
\usepackage{xcolor}
\newcommand{\D}[1]{\begin{normalsize}\begin{tt}#1\end{tt}\end{normalsize}}

\newcommand\StateX{\Statex\hspace{\algorithmicindent}}

\newcommand{\rw}[1]{\textcolor{black}{#1}}

\definecolor{purple_}{HTML}{7E57C2}
\definecolor{orange_}{HTML}{F57C00}

\articletype{Research Article}%

\received{23 June 2021}
\revised{15 October 2021}
\accepted{26 December 2021}
        
\begin{document}


\title{Towards Fast Theta-join: A Pre-filtering and Amalgamated Partitioning Approach}

\author[1,2]{Jiashu Wu}
\author[1]{Yang Wang*}
\author[1]{Xiaopeng Fan}
\author[1]{Kejiang Ye}
\author[3]{Chengzhong Xu}

\authormark{Jiashu Wu \textsc{et al}}

\address[1]{\orgdiv{Shenzhen Institute of Advanced Technology}, \orgname{Chinese Academy of Sciences}, \orgaddress{\state{Shenzhen 518055}, \country{China}}}
\address[2]{\orgname{University of Chinese Academy of Sciences}, \orgaddress{\state{Beijing 100049}, \country{China}}}
\address[3]{\orgname{University of Macau}, \orgaddress{\state{Macau 999078}, \country{China}}}

\corres{*Corresponding author: Yang Wang, \email{yang.wang1@siat.ac.cn}}


\abstract[Abstract]{As one of the most useful online processing techniques, the theta-join operation has been utilised by many applications to fully excavate the relationships between data streams in various scenarios. As such, constant research efforts have been put to optimise its performance in the distributed environment, which is typically characterised by reducing the number of Cartesian products as much as possible. 
In this paper, we design and implement a novel fast theta-join algorithm, called \emph{Prefap}, by developing two distinct techniques---\emph{\underline{PRE}-\underline{F}iltering} and \emph{\underline{A}malgamated \underline{P}artitioning}---based on the state-of-the-art FastThetaJoin algorithm to optimise the efficiency of the theta-join operation. Firstly, we develop a pre-filtering strategy before data streams are partitioned to reduce the amount of data to be involved and benefit a more fine-grained partitioning. Secondly, to avoid the data streams being partitioned in a coarse-grained isolated manner and improve the quality of the partition-level filtering, we introduce an amalgamated partitioning mechanism that can amalgamate the partitioning boundaries of two data streams to assist a fine-grained partitioning. With the integration of these two techniques into the existing FastThetaJoin algorithm, we design and implement a new framework to achieve a decreased number of Cartesian products and a higher theta-join efficiency. By comparing with existing algorithms, FastThetaJoin in particular, we evaluate the performance of \emph{Prefap} on both synthetic and real data streams from two-way to multi-way theta-join to demonstrate its superiority. }

\keywords{Theta-join ($\theta$-join), Online data stream, Pre-filtering, Amalgamated data stream partitioning, Cartesian product reduction}

\maketitle


\section{Introduction}\label{sec:sec1_introduction}

As the big data technology becomes more prevalent \cite{big_data_more_prevalent_buckee2020improving,big_data_more_prevalent_tsai2015big} and widely deployed \cite{big_data_widely_deployed_deng2020big,big_data_widely_deployed_wang2020big,big_data_widely_deployed_marjani2017big,big_data_widely_deployed_luo2016big}, tremendous amount of online data streams have been generated \cite{online_data_stream_generated_rapidly_din2020online,online_data_stream_generated_rapidly_dong2020interactive,online_data_stream_generated_rapidly_bifet2015efficient}. In the financial market \cite{stock_price_data_stream_rezaei2021stock,stock_price_data_stream_silva2013data}, the price of stocks keeps fluctuating, the currency conversion rates change every few seconds in an online manner. For the meteorological monitoring services \cite{cloud_dataset_reference_hahn1999extended}, thousands of monitoring stations constantly monitor the meteorological data in real-time \cite{metero_data_stream_vaananen2020sensor}, such as wind speed and temperature, etc. Therefore, how to process these online data streams efficiently and fully excavate the knowledge \cite{data_stream_knowledge_gaber2005mining,data_stream_knowledge_gama2010knowledge} behind them become crucial to explore. 

\begin{figure}[t]
  \begin{center}
    \includegraphics[width=0.6\textwidth,keepaspectratio]{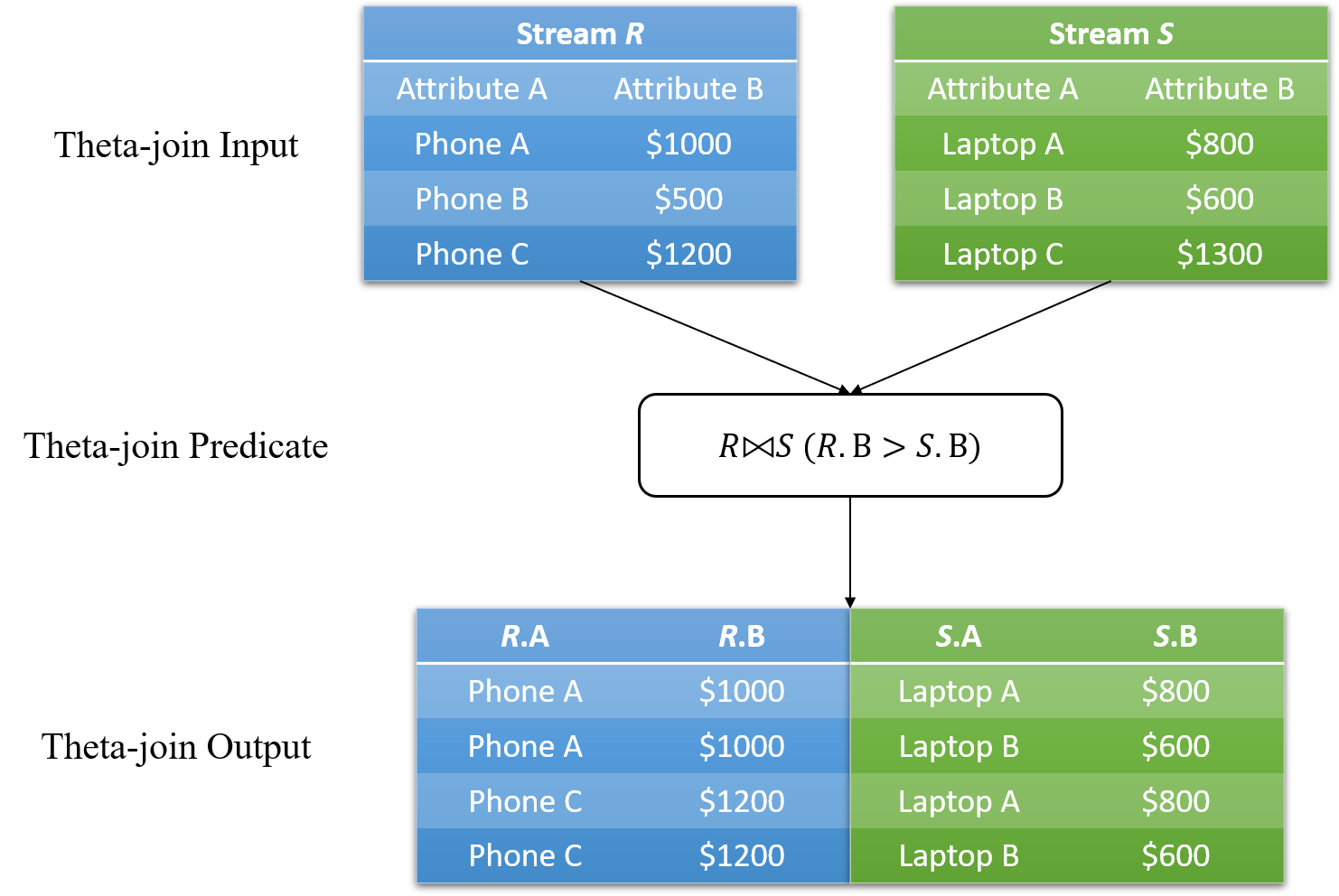}\\
    \caption{An illustrative example of a theta-join predicate $\mathcal{R}$ $\bowtie$ $\mathcal{S}$ ($\mathcal{R}$.B $>$ $\mathcal{S}$.B), its input and its corresponding output. In this example, there are only 4 phone-laptop combinations that satisfy the price of the phone is more expensive than laptop. }
    \label{fig:figure0}
  \end{center}
\end{figure}

To relate data streams together, the join operation \cite{join_operation_garcia2008database,cartesian_product_time_consuming_mishra1992join} is one of the vital operations that is capable to detect scenarios that satisfy certain conditions. To find out data elements in two data streams that are equal, equi-join \cite{equi_join_lee2012join,equi_join_rui2017fast,equi_join_chen2016accelerating} should be used. As for non-equal relationships between data streams, the theta-join \cite{theta_join_wilschut1990pipelining,theta_join_penar2015evaluation} comes to help. Theta-join, denoted as follows 

\begin{equation}
  \label{equ:equation_two_way_theta_join_formulation}
  \mathcal{R} \bowtie \mathcal{S} (\text{A} \ \theta \ \text{B})
\end{equation}
serves as a special kind of join operation that relates the attribute A of the data stream $\mathcal{R}$ and the attribute B of the data stream $\mathcal{S}$ using a non-equal theta condition among one of the following $\{<, \leq, >, \geq\}$. Cartesian products \cite{cartesian_product_theta_join_koumarelas2014binary,cartesian_product_theta_join_bellas2017gpu} between two data streams will be generated and data element pairs that satisfy the theta condition are picked out to form the theta-join results. For instance, as illustrated in Figure \ref{fig:figure0}, the inputs of this two-way theta-join operation are data stream Phone (Stream $\mathcal{R}$) and data stream Laptop (Stream $\mathcal{S}$). Under the join predicate $\mathcal{R}$ $\bowtie$ $\mathcal{S}$ ($\mathcal{R}$.B $>$ $\mathcal{S}$.B), all phone-laptop combinations that satisfy the price of the phone is higher than the price of the laptop will be returned as results. Therefore, 4 satisfied combinations are returned as output as shown in Figure \ref{fig:figure0}. 

Moreover, the theta-join can also be generalised to work on multiple data streams. For instance, the 3-way theta-join predicate is expressed as follows

\begin{equation}
  \label{equ:equation_multi_way_theta_join_formulation}
  \mathcal{R} \bowtie \mathcal{S} \bowtie \mathcal{T} (\text{A} \ \theta_1 \ \text{B} \ \theta_2 \ \text{C})
\end{equation}
which exploits all A, B, C attribute combinations that satisfy the A $\theta_1$ B $\theta_2$ C condition from three data streams $\mathcal{R}$, $\mathcal{S}$ and $\mathcal{T}$, respectively. When dealing with data streams that come in an online manner, the theta-join algorithm will receive the data stream in the form of a window, with the window size $w$ denotes the amount of data consisted in the window. Then, the theta-join operation will be performed between windows of data streams. As a powerful data analysis and processing tool \cite{powerful_data_processing_tool_huge_computing_overhead_cao2018optimization}, theta-join can be widely utilised in broad applications, such as discerning on which day stock $\mathcal{R}$ performs worse than stock $\mathcal{S}$ in 2019, or whether the wind speed in July is faster than June 2020 in most of the days. \rw{The theta-join operation has also been implemented in popular database management systems such as Oracle database \cite{theta_join_oraclesql}, PostgreSql database \cite{theta_join_postgresql}, etc. }

However, the theta-join's efficiency is heavily affected by the number of involved time-consuming Cartesian products \cite{cartesian_product_time_consuming_mishra1992join,cartesian_product_time_consuming_sviridov2014performance}. Handling the theta-join in a careless way can result in the number of Cartesian products grows drastically, and will even lead to an exponential surge if multi-way theta-join is conducted. The unacceptably large number of Cartesian products also become a curse, especially in distributed environments \cite{large_data_overhead_del2020rhino,large_data_overhead_rodrigues2009multi,large_data_overhead_ur2016big} where the incurred I/O overhead and the communication cost \cite{powerful_data_processing_tool_huge_computing_overhead_cao2018optimization} make it an influential factor that dramatically compromises the efficiency of the theta-join operation. Therefore, they should be reduced as much as possible. As a result, numerous efforts have been made by both industry and academia from different aspects to optimise the efficiency of the theta-join operation \cite{range_based_method_dewitt1992practical,one_bucket_theta_okcan2011processing,cross_filter_strategy_liu2017efficient,fastthetajoin_original_hu2020fastthetajoin}. 


Although the existing algorithms in the literature exhibited different merits in making the theta-join operation more efficient, they still suffered from some deficiencies in effective Cartesian product reduction and data stream partitioning, which more or less compromised the theta-join operation efficacy. Therefore, we still have rooms to further improve them in various ways. In this paper, we propose a novel fast online theta-join algorithm, called \emph{Prefap}. Instead of performing filtering only after partitioning as in \cite{range_based_method_dewitt1992practical,one_bucket_theta_okcan2011processing,cross_filter_strategy_liu2017efficient,fastthetajoin_original_hu2020fastthetajoin}, it makes sense to perform a \emph{pre-filtering} 
based on the theta condition before the partitioning takes place as it can not only reduce the amount of data involved in the partitioning but also make the partitioning more fine-grained. Since our partitioning is based on the range boundaries calculated using the minimum and the maximum values of the data streams, the pre-filtering of the data streams is possible to condense the range between the minimum and the maximum value, hence making the partition more fine-grained. Moreover, this pre-filtering mechanism does not suffer from the severe overhead caused by the time-consuming sorting operation \cite{sorting_time_consuming_aliyu2013comparative,sorting_time_consuming_al2013review}. As opposed to \cite{range_based_method_dewitt1992practical,one_bucket_theta_okcan2011processing,cross_filter_strategy_liu2017efficient}, the proposed algorithm requires neither inter-partition nor intra-partition ordering. 

Furthermore, for all the aforementioned algorithms, the partitioning of two data streams are conducted in a coarse-grained isolated manner, i.e., the partitioning of data stream $\mathcal{R}$ has no impact on the partitioning of data stream $\mathcal{S}$, which may impair the effectiveness of the filtering mechanism after the partitioning. 
To overcome the drawback incurred by the isolated coarse-grained partitioning, the proposed algorithm introduces an \emph{amalgamated partitioning} mechanism. With this mechanism, the data stream will be partitioned based on the amalgamated partitioning boundary that fuses the partitioning information of data streams. Hence, worthless Cartesian products between partitions that are deemed impossible to possess valid theta-join results will be avoided, and hence making the algorithm more efficient, which is an improvement to~\cite{range_based_method_dewitt1992practical,one_bucket_theta_okcan2011processing,cross_filter_strategy_liu2017efficient,fastthetajoin_original_hu2020fastthetajoin}. 

To validate the effectiveness of the proposed method and to make it applicable, we integrate the pre-filtering strategy and the amalgamated partitioning mechanism to form a new theta-join processing framework. The framework uses FastThetaJoin~\cite{fastthetajoin_original_hu2020fastthetajoin} as a basis. However, the proposed \emph{Prefap} framework makes contributions by performing pre-filtering and avoiding the isolated partitioning. With the \emph{Prefap} framework, we can substantially boost the efficiency of the theta-join operation. The proposed \emph{Prefap} framework is comprehensively evaluated on both synthetic and real data streams in distributed environments and compared with other algorithms, demonstrating its superior performance. 

Therefore, in summary, the paper makes the following contributions: 

\begin{itemize}
  \item We propose a pre-filtering strategy that can not only reduce the amount of data involved in the partitioning but also make the partitioning more fine-grained. 
  \item We introduce the amalgamated partitioning mechanism that avoids the coarse-grained isolated partitioning and hence benefits the reduction of Cartesian products and makes the algorithm more efficient. 
  \item We unify the proposed pre-filtering strategy and the amalgamated partitioning mechanism to form a holistic framework, called \emph{Prefap}, that can boost the effectiveness of theta-join operations while avoiding the time-consuming burdens such as sorting. 
  \item The \emph{Prefap} framework is implemented and comprehensive empirical evaluations on both synthetic and real data streams are conducted to testify the superiority. 
\end{itemize}

The remainders of the paper are organised as follows. Section \ref{sec:sec2_relatedwork} overviews some related works to show the distinct features of the proposed algorithm. The \emph{Prefap} algorithm is introduced in Section \ref{sec:sec3_the_proposed_algorithm}. Section \ref{sec:sec4_experiments_and_results_analysis} presents the experimental results, which are also comprehensively analysed and discussed. The last section concludes the paper. 

\begin{figure}[t]
  \begin{center}
    \includegraphics[height=8.1cm,keepaspectratio]{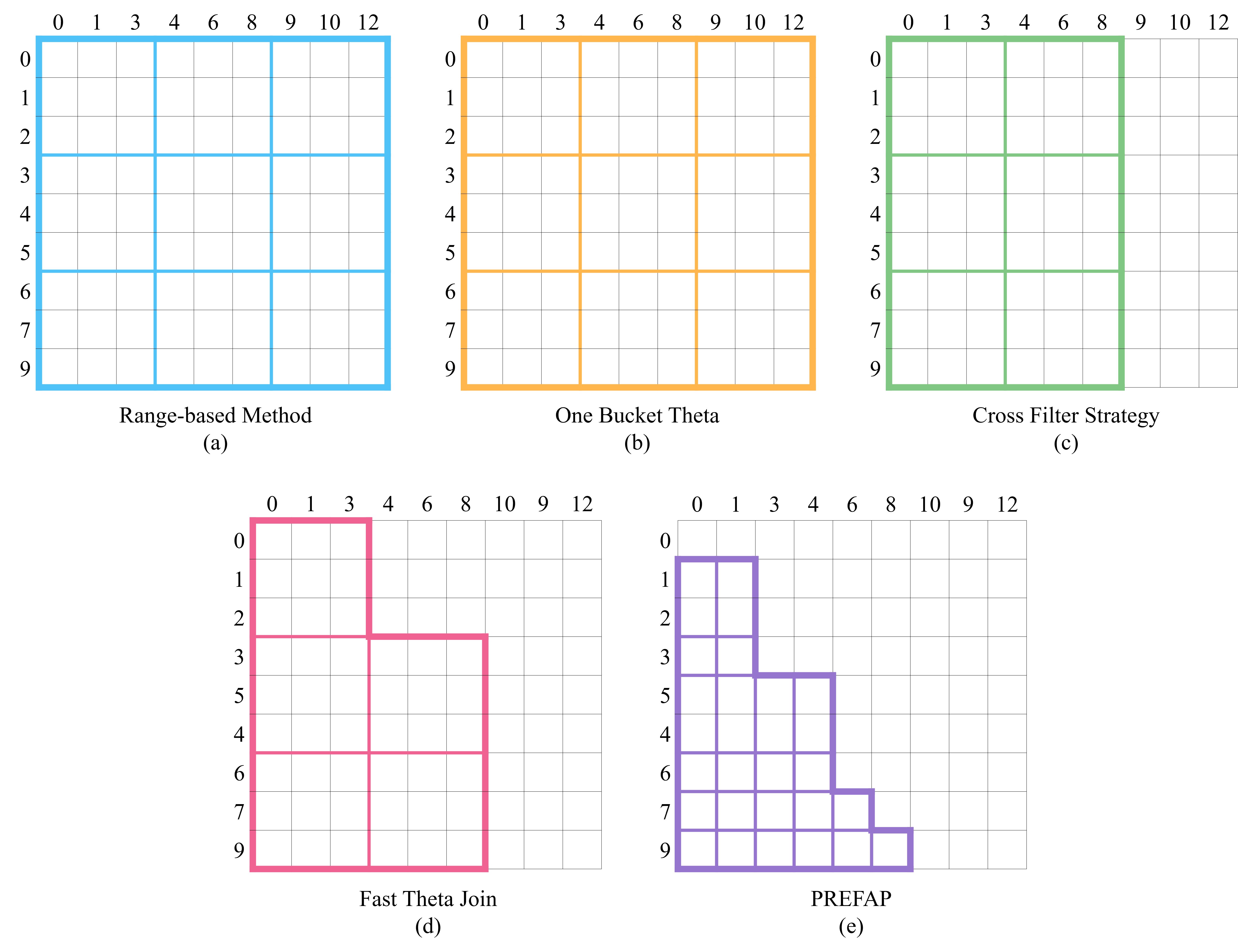}\\
    \caption{\rw{Illustration of the number of Cartesian products involved in each algorithm. The theta condition in this example is $\mathcal{R} > \mathcal{S}$. The data stream $\mathcal{R}$ is placed vertically and the data stream $\mathcal{S}$ is placed horizontally. The thick coloured border indicates the boundary of Cartesian products that need to be performed, while the thin coloured border indicates the partitioning boundary. }}
    \label{fig:figure1}
  \end{center}
\end{figure}

\begin{figure}[!h]
  \begin{center}
    \includegraphics[height=5cm,keepaspectratio]{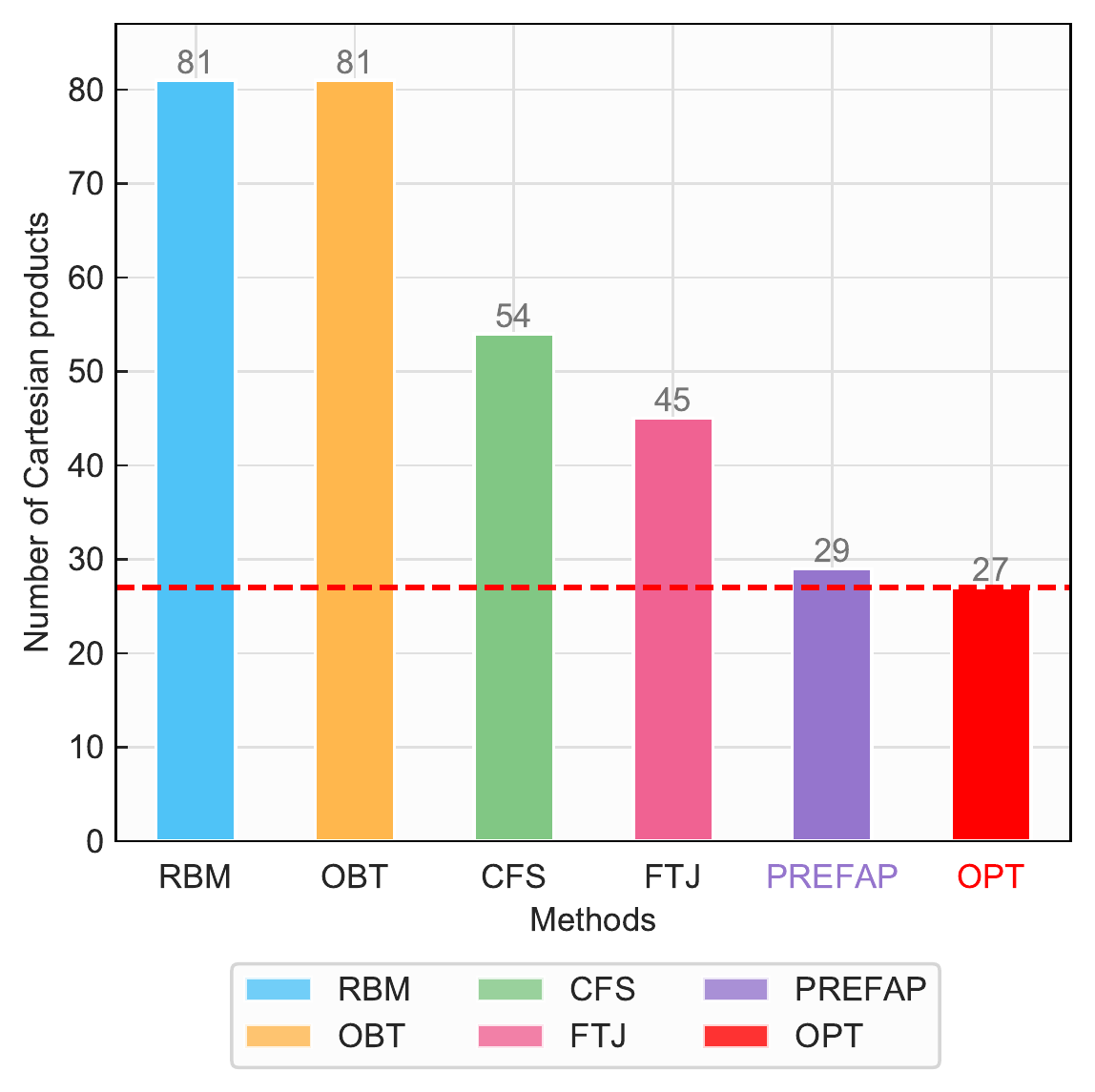}\\
    \caption{\rw{The bar chart presents the number of Cartesian products performed by each algorithm in the example in Figure \ref{fig:figure1}. The rightmost red bar, as well as the red dotted line, indicate the number of theta-join results, i.e., no matter how the theta-join algorithm is optimised, this is the lower bound of the number of Cartesian products that the algorithm needs to perform and the algorithm cannot perform better than this. }}
    \label{fig:figure_auxiliary}
  \end{center}
\end{figure}


\section{\rw{Related Work and Research Opportunity}}
\label{sec:sec2_relatedwork}

As the scale of data streams keeps growing rapidly \cite{online_data_stream_generated_rapidly_din2020online,online_data_stream_generated_rapidly_dong2020interactive,online_data_stream_generated_rapidly_bifet2015efficient}, how to efficiently utilise the theta-join to process data streams becomes a crucial problem, and thereby attracting numerous attentions from both industry and academic communities. In this section, we first overview some past research efforts on improving the efficiency of the theta-join operation, and then identify their deficiencies to show our research opportunities. 

\subsection{Range-based Method}\label{sec:sec2.1_range_based_partitioning_method}
Dewitt et al., \cite{range_based_method_dewitt1992practical} proposed a Range-based Method (RBM) to carry out the theta-join operation in distributed environments. The joint attributes in data streams were firstly sorted, then two data streams were partitioned where the partitioning boundaries were characterised by sampled ranges, and Cartesian products were performed without any filtering being performed. 

For example, in Figure \ref{fig:figure1}(a), if we set the number of partitions to $3$, the Range-based Method will sample the 4\textsuperscript{th} and the 7\textsuperscript{th} value from the sorted data stream $\mathcal{R}$, which are $3$ and $6$, respectively. Then the data stream $\mathcal{R}$ is partitioned based on the ranges as follows: \D{($-\infty$,3)}, \D{[3,6)} and \D{[6,$+\infty$)}. The same partitioning is applied for data stream $\mathcal{S}$ as well. 

However, the Range-based Method suffered from several disadvantages, which made it extremely time-consuming. Firstly, sorting data streams is very inefficient \cite{sorting_time_consuming_aliyu2013comparative,sorting_time_consuming_al2013review}, especially when the size of the data stream is large. Secondly, the Range-based Method does not apply any kind of filtering, hence, entire data streams will participate in the Cartesian products, resulting in a huge number of Cartesian products being performed. Finally, the Range-based Method suffers a lot from data skewness. For instance, \D{[1,1,1,1,1,1,2,2,3]} will be partitioned into \D{($-\infty$,1)}, \D{[1,2)} and \D{[2,$+\infty$)}, which could lead to severely imbalanced workloads between partitions, and is especially hurtful in distributed environments. 

\subsection{Randomised Method}\label{sec:sec2.2_randomised_method}

To address the issues faced by the Range-based Method, Okcan et al., \cite{one_bucket_theta_okcan2011processing} put forward an algorithm, called One-Bucket-Theta (OBT), which partitioned the data stream as evenly as possible, and then randomly distributed the partitioned blocks in the distributed environment. 

For instance, in Figure \ref{fig:figure1}(b), the One-Bucket-Theta algorithm first sorts the data streams. Then, take the number of partitions to be $3$ as an example, instead of partitioning the data stream based on ranges, the One-Bucket-Theta partitions the $1$\textsuperscript{st} $\rightarrow$ $3$\textsuperscript{rd}, $4$\textsuperscript{th} $\rightarrow$ $6$\textsuperscript{th}, and $7$\textsuperscript{th}  $\rightarrow$ $9$\textsuperscript{th} data elements into $3$ partitions, respectively, resulting in a relatively more even partitioning scheme in practice. 

Furthermore, the randomised distribution of partitions to processes in the distributed environment further balances the workloads among processes and avoids the severe load imbalance. However, the One-Bucket-Theta still suffers from the time complexity brought by its sorting operation. Also, the lack of filtering strategy makes entire data streams being involved when performing the Cartesian products, which significantly impairs its efficiency. 

\subsection{Filtering Method}\label{sec:sec2.3_filtering_method}

In order to improve the efficiency of the theta-join by reducing the number of Cartesian products, the Cross Filter Strategy (CFS) proposed by Liu et al., \cite{cross_filter_strategy_liu2017efficient} performed stream-level filtering after the data stream partitioning to eliminate data elements that are unlikely to form valid theta-join results based on the theta condition. 

As shown in Figure \ref{fig:figure1}(c), under the ``$>$'' theta condition, the Cartesian products will not be performed between the entire data stream $\mathcal{R}$ and partition \D{[9,12]} of data stream $\mathcal{S}$. Since the maximum value $9$ of data stream $\mathcal{R}$ equals to the minimum value $9$ in partition \D{[9,12]} of data stream $\mathcal{S}$, hence there is no way for the Cartesian products between data stream $\mathcal{R}$ and range \D{[9,12]} of data stream $\mathcal{S}$ to possess valid ``$>$'' theta-join results, and hence partition \D{[9,12]} of data stream $\mathcal{S}$ is eliminated by the stream-level filtering strategy. Following the same principle, the stream-level filtering can also be applied to other similar theta conditions as in $\{\ge, <, \le\}$. 

However, this stream-level filtering is coarse-grained and fails to remove unnecessary Cartesian products as much as possible. The stream-level filtering finds out that partition \D{[0,3)} of data stream $\mathcal{R}$ can form valid theta-join results with partition \D{[0,4)} of data stream $\mathcal{S}$ under the ``$>$'' theta condition, and hence partition \D{[0,3)} of data stream $\mathcal{R}$ could not be eliminated and the Cartesian products between it and the entire filtered data stream $\mathcal{S}$ will be performed, even though the Cartesian products between \D{[0,3)} and \D{[4,9)} are totally redundant. 

To further improve the efficiency of the theta-join operation in terms of reducing the number of Cartesian products, Hu et al., \cite{fastthetajoin_original_hu2020fastthetajoin} presented the FastThetaJoin (FTJ) algorithm. In contrast to performing filtering at the stream-level, the FastThetaJoin algorithm utilised the partition-level filtering strategy, which compared all the partition pairs from both data streams and avoided performing the Cartesian products between the partitions that are unable to form any valid theta-join results based on the theta condition as shown in Figure \ref{fig:figure1}(d). However, the lack of the pre-filtering mechanism before partitioning not only incurred all data elements in two data streams to be involved in partitioning, irrespective of whether they are able to form valid theta-join results or not, but also 
made the partitioning more coarse-grained, and hence compromised the overall performance. 

Furthermore, in the FastThetaJoin algorithm, the partitioning of data streams was conducted in an isolated manner. Failing to partition the data streams collaboratively made the FastThetaJoin highly laborious as the coarse-grained isolated data stream partitioning is incapable to remove some worthless data elements in some partitions. In terms of auxiliary procedures, the FastThetaJoin algorithm adopted the re-partitioning on oversized partitions so that the workloads would be well-balanced and the method will be more distributed-environment-friendly. The Cartesian products were performed between the remaining partitions of two data streams. 

\subsection{\rw{Research Opportunity}}
\label{sec:sec2.4_motivation}

Despite constant optimisation efforts being made, there is still room to extend and improve the efficiency of the theta-join operation, hence it brings us the motivation of the proposed framework. As shown in Figure \ref{fig:figure1}(d), performing the filtering strategy after partitioning ends up with performing the Cartesian products between $0$ in data stream $\mathcal{R}$ and the partition \D{[0,4)} of data stream $\mathcal{S}$, which is unnecessary under the ``$>$'' theta condition. Hence, a pre-filtering strategy that filters the data streams before the partitioning will be beneficial to reduce the amount of data that needs to be partitioned, and it can also make the partitioning become more fine-grained as the pre-filtering shortens the ranges of the data streams. 

Moreover, the isolated partitioning of data streams can be substituted with the amalgamated partitioning mechanism. Under the isolated manner with the theta condition as ``$>$'', the Cartesian products between partition \D{[3,6)} and \D{[4,8)} will be conducted simply because of a single valid theta-join result which is $5 > 4$, but the Cartesian products between \D{[3,4]} and \D{[4,8)} are completely unnecessary. To avoid these pointless Cartesian products incurred by the isolated partitioning, the partitioning information of two data streams will be amalgamated to form an amalgamated partitioning scheme so that the partition \D{[3,6)} will be split to filter more unnecessary Cartesian products that possess no valid theta-join result. 

Hence, by introducing the pre-filtering strategy and the amalgamated partitioning mechanism, we can integrate them into a unified framework to effectively reduce the amount of unnecessary Cartesian products while balancing the workload in the distributed environment and avoiding the time-consuming sorting operation. Therefore, we can substantially boost the efficiency of the theta-join operation. 

\rw{As illustrated in Figure \ref{fig:figure_auxiliary}, the bar chart presents the number of Cartesian products that are performed by each algorithm in the example in Figure \ref{fig:figure1}. The rightmost red bar, as well as the red dotted line, indicate the number of theta-join results, i.e., no matter how the theta-join algorithm is optimised, it is the lower bound of the number of Cartesian products that needs to be performed. The proposed \emph{Prefap} algorithm achieves the lowest number of Cartesian products in this example and is very close to the optimal lower bound, which indicates that the research direction of the \emph{Prefap} algorithm is promising. }


\begin{figure}[!h]
  \begin{center}
    \includegraphics[height=8.5cm,keepaspectratio]{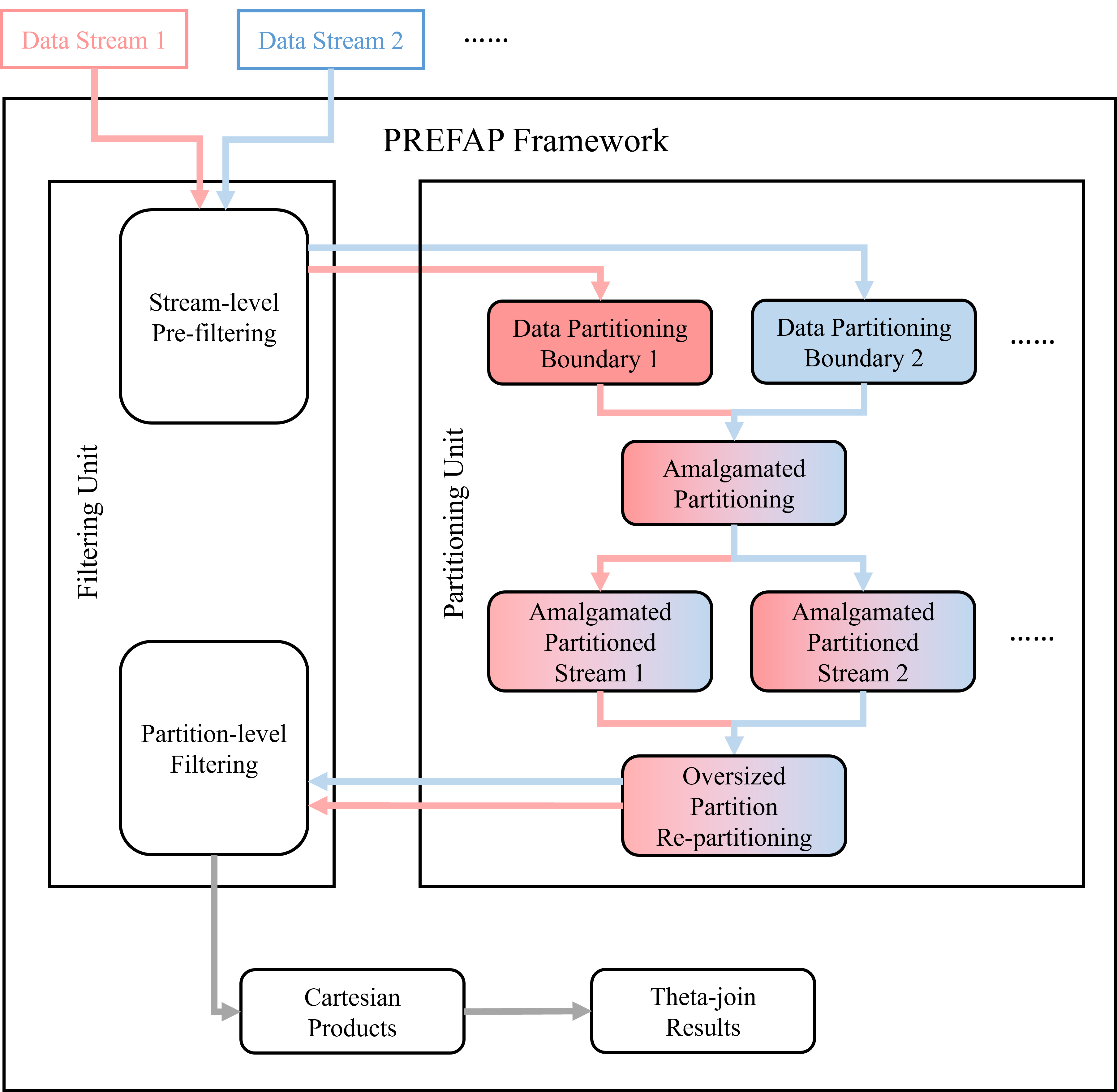}\\
    \caption{\rw{The \emph{Prefap} framework. In the Filtering Unit of the \emph{Prefap} framework, the data streams will firstly be pre-filtered based on the theta condition to get rid of unnecessary data elements that are deemed not possible to form any valid theta-join results. Then, their corresponding partitioning boundaries will be calculated by the Partitioning Unit, followed by the amalgamated partitioning procedure, in which the partitioning boundaries are amalgamated. The resulted data streams are partitioned based on the amalgamated partitioning boundaries, and those oversized partitions are re-partitioned to achieve load balancing in the distributed environment. After that, the partitions will be handled by the Filtering Unit and the partition-level filtering is performed. Finally, Cartesian products are conducted and the theta-join results are retrieved based on the theta condition. Note that the diagram shows how two data streams are processed by the \emph{Prefap} framework for the purpose of illustration only, the framework can be extended to work on multi-way data stream theta-join as well. For three-way theta-join, two data streams will firstly be processed by the \emph{Prefap} framework, the result produced will then be joined with the third data stream to yield the final result. }}
    \label{fig:figure2}
  \end{center}
\end{figure}

\section{Prefap Framework}\label{sec:sec3_the_proposed_algorithm}

In this section, we will introduce the proposed \emph{Prefap} approach in terms of its framework and workflow, followed by a detailed explanation of each constituting component, i.e., the pre-filtering strategy, the amalgamated partitioning scheme, as well as some auxiliary steps to complete the theta-join operation. Note that in this section and the corresponding pseudocode, we explain the theta-join operation on two data streams for the sake of simplicity. However, it is not difficult to extend our approach to multi-way theta-joins. The symbols and acronyms used and their corresponding interpretations are given in Table \ref{tab:table_symbol_description}. 

\subsection{Prefap Workflow}
\label{sec:sec3.1_the_workflow_of_the_proposed_framework}

The framework of the proposed \emph{Prefap} is illustrated in Figure \ref{fig:figure2}, together with its workflow as follows: 

\textit{Step 1 - Pre-filtering Strategy}: \rw{Firstly, the proposed pre-filtering strategy will be employed to filter out data elements in two data streams based on the theta condition, so that the workloads in later steps can be lowered and the partitioning will be more fine-grained (as in Section \ref{sec:sec3.2_the_pre_filtering_strategy} and Algorithm \ref{alg:alg_step1}). This step will be completed by the Filtering Unit of the \emph{Prefap} framework. }

\textit{Step 2 - Amalgamated Partitioning Mechanism}: \rw{Then, it comes to the amalgamated partitioning mechanism. The partitioning boundaries of data streams are calculated, amalgamated, and the amalgamated partitioning boundaries will be used to produce fine-grained partitions to benefit the partition-level filtering (as in Section \ref{sec:sec3.3_amalgamated_partitioning} and Algorithm \ref{alg:alg_step2}). This step will be completed by the Partitioning Unit of the \emph{Prefap} framework. }

\textit{Step 3 - Auxiliary  Procedures:} Finally, after completing the developed pre-filtering strategy and the amalgamated partitioning mechanism, some auxiliary procedures adopted in FastThetaJoin are followed to complete the theta-join operation for the final output results: 

\begin{enumerate}
\item \textit{Step 3.1 - Oversized Partition Re-partitioning}: \rw{To balance the workload among processes under the distributed environment, oversized partitions will be re-partitioned to balance the workload as much as possible. This step will be completed by the Partitioning Unit of the \emph{Prefap} framework. }

\item \textit{Step 3.2 - Partition-level Filtering}: \rw{The partitions will then be filtered again based on the theta condition to avoid unnecessary Cartesian products. This step will be completed by the Filtering Unit of the \emph{Prefap} framework. }

\item \textit{Step 3.3 - Cartesian Products and Theta-join Results}: Finally, the Cartesian products are performed and the theta-join results are retrieved based on the theta condition. All auxiliary procedures will be presented in Section \ref{sec:sec3.4_auxiliary_procedures} and Algorithm \ref{alg:alg_step3}. 
\end{enumerate}

\begin{table}[!ht]
  \caption{\rw{Interpretation of symbols and acronyms. Input parameters are marked with $(\star)$}}
  \centering
  \resizebox{0.7\textwidth}{!}{\begin{tabular}{c|c}
  \hline
  \bfseries \rw{Symbol / acronym} & \bfseries \rw{Interpretation}\\
  \hline\hline
  \rw{$\uptheta$ $(\star)$} & \rw{The $\uptheta$ operator, $\uptheta \in \{<, \leq, >, \geq\}$}\\
  \hline
  \rw{$\mathcal{R, S, T}$ $(\star)$} & \rw{Data stream R, S, T}\\
  \hline
  \rw{$D$} & \rw{A variable which stands for data stream, $D \in \{\mathcal{R}, \mathcal{S}, \mathcal{T}, \dots\}$}\\
  \hline
  \rw{$D_{min}$} & \rw{The minimum value of entire data stream $D$}\\
  \hline
  \rw{$sp_D$} & \rw{The span of each partition of data stream $D$}\\
  \hline
  \rw{$p$ $(\star)$} & \rw{Number of partitions}\\
  \hline
  \rw{$PB_D$} & \rw{Partitioning boundary of data stream $D$ before being amalgamated}\\
  \hline
  \rw{$APB$} & \rw{Amalgamated partitioning boundary}\\
  \hline
  \rw{$P_D$} & \rw{Partitions of data stream $D$}\\
  \hline
  \rw{$P^i_D$} & \rw{The $i$\textsuperscript{th} partition of data stream $D$}\\
  \hline
  \rw{$rn^i_D$} & \rw{The re-partitioning number of $P^i_D$}\\
  \hline
  \rw{$AS_D$} & \rw{The average partition size of data stream $D$}\\
  \hline
  \rw{$sp^{i'}_D$} & \rw{The re-partitioning span of the $i$\textsuperscript{th} partition of data stream $D$}\\
  \hline
  \rw{$w$ $(\star)$} & \rw{The window size}\\
  \hline

  \end{tabular}}
  \label{tab:table_symbol_description}
\end{table}

\begin{algorithm*}[!ht]
  \begin{algorithmic}[1]
    \Require
      \StateX $\uptheta$ operator, $\uptheta$ $\in$ \{$<, \leq, >, \geq$\}, 
      \StateX Data streams $\mathcal{R}$ and $\mathcal{S}$, attribute $\mathcal{R}$.A and $\mathcal{S}$.B
    \Ensure Pre-filtered data streams at the stream-level. 
    \If{$\theta$ \textup{is} ``$>$''}
      \State $\mathcal{R}$ $\leftarrow$ $\mathcal{R}$.remove($\leq \mathcal{S}_{min}$)
      \State $\mathcal{S}$ $\leftarrow$ $\mathcal{S}$.remove($\geq \mathcal{R}_{max}$)
    \ElsIf{$\theta$ \textup{is} ``$\geq$''}
      \State $\mathcal{R}$ $\leftarrow$ $\mathcal{R}$.remove($< \mathcal{S}_{min}$)
      \State $\mathcal{S}$ $\leftarrow$ $\mathcal{S}$.remove($> \mathcal{R}_{max}$)
    \ElsIf{$\theta$ \textup{is} ``$<$''}
      \State $\mathcal{R}$ $\leftarrow$ $\mathcal{R}$.remove($\geq \mathcal{S}_{max}$)
      \State $\mathcal{S}$ $\leftarrow$ $\mathcal{S}$.remove($\leq \mathcal{R}_{min}$)
    \Else
      \State $\mathcal{R}$ $\leftarrow$ $\mathcal{R}$.remove($> \mathcal{S}_{max}$)
      \State $\mathcal{S}$ $\leftarrow$ $\mathcal{S}$.remove($< \mathcal{R}_{min}$)
    \EndIf
    \State \Return $\mathcal{R}, \mathcal{S}$
  \end{algorithmic}
  \caption{\rw{The \emph{Prefap} algorithm - Stream-level Pre-filtering Strategy (Step 1)}}
  \label{alg:alg_step1}
\end{algorithm*}

\subsection{Pre-filtering Strategy}
\label{sec:sec3.2_the_pre_filtering_strategy}

\rw{As opposed to all the aforementioned algorithms that directly perform the partitioning without any pre-filtering, a pre-filtering strategy is applied in the \emph{Prefap} framework to eliminate certain amounts of unnecessary data involved in the operation. This is performed in the Partitioning Unit and the the pseudocode of this step is given in Algorithm \ref{alg:alg_step1}. }Specifically, the pre-filtering strategy scans and filters the two data streams according to the theta condition to eliminate the data elements that are deemed impossible to produce valid theta-join results before the partitioning is performed. Also, unlike some previous methods, the proposed pre-filtering strategy does not require data stream sorting, which severely hurts the efficiency of the theta-join operation. \rw{For example, as illustrated in Figure \ref{fig:figure1}(e), given the theta condition is ``$>$'', any value in data stream $\mathcal{R}$ that is less than or equal to the minimum value of data stream $\mathcal{S}$ is safe to be removed as it is not possible to form valid theta-join results with any value in data stream $\mathcal{S}$. The similar mechanism applies for data stream $\mathcal{S}$ as well. Therefore, we can safely eliminate any value in data stream $\mathcal{S}$ that is greater than or equal to the maximum value of data stream $\mathcal{R}$, as is shown in line $2$-$3$ in Algorithm \ref{alg:alg_step1}. The pre-filtering mechanism works similarly for other theta conditions as presented in Algorithm \ref{alg:alg_step1}. As indicated in Figure \ref{fig:figure1}(e), upon applying the pre-filtering strategy, the first row and the last three columns will be directly filtered out and not involved in later processing steps. Hence the pre-filtering strategy can greatly reduce the amount of data that needs to be processed and improve the efficiency of the theta-join operation. }

Meanwhile, since the partitioning boundaries being used in the subsequent processing are calculated based on the minimum and maximum values of data streams, the use of the pre-filtering strategy to filter useless data is likely to reduce the span between the maximum and the minimum values, and therefore making the partitioning more fine-grained and benefiting the partition-level filtering performed later. 

\begin{algorithm*}[!ht]
  \begin{algorithmic}[1]
    \Require
      \StateX $\uptheta$ operator, $\uptheta$ $\in$ \{$<, \leq, >, \geq$\}, 
      \StateX number of partitions $p$, 
      \StateX Data streams $D$ after stream-level pre-filtering, $D \in \{\mathcal{R}, \mathcal{S}\}$
    \Ensure Amalgamated partitioned data streams
    \State // Amalgamated Partitioning Mechanism
    \State Calculate $sp_D$ based on Equation (\ref{equ:equation1_partition_span_calculation})
    \State $PB_D$ $\leftarrow$ [ $D_{min}$, $D_{min} + 1 \times sp_D$ ), 
    \StateX \hspace*{1.7em} [ $D_{min} + 1 \times sp_D$, $D_{min} + 2 \times sp_D$ ), 
    \StateX \hspace*{7.6em} \vdots
    \StateX \hspace*{1.7em} [ $D_{min} + (p - 1) \times sp_D$, $D_{max}$ ]
    \State // Generate amalgamated partitioning boundary
    \State $APB$ $\leftarrow$ $PB_\mathcal{R}$.append($PB_\mathcal{S}$).sort()
    \State // Partition data streams using the amalgamated partitioning boundary
    \State $P_D$ $\leftarrow$ Partition $D$ based on $APB$
    \State $P_D$ $\leftarrow$ $P_D$.filter($P_D^i$.size() != 0)
    \State \Return $P_D$, $D \in \{\mathcal{R}, \mathcal{S}\}$
  \end{algorithmic}
  \caption{\rw{The \emph{Prefap} algorithm - Amalgamated Partitioning Mechanism (Step 2)}}
  \label{alg:alg_step2}
\end{algorithm*}

\subsection{Amalgamated Partitioning Mechanism}
\label{sec:sec3.3_amalgamated_partitioning}

\rw{The data streams are partitioned in the Partitioning Unit based on the range defined by the partitioning boundaries in the course of the theta-join operation as shown in line $3$ in Algorithm \ref{alg:alg_step2}. The partitioning boundaries of data streams will be calculated after the pre-filtering strategy is executed with each partition having a span, which is denoted as \emph{sp} and is calculated as follows: }

\rw{\begin{equation}
  \label{equ:equation1_partition_span_calculation}
  sp_D = \frac{D_{max} - D_{min}}{p}, D \in \{\mathcal{R}, \mathcal{S}\}
\end{equation}}
where $p$ denotes the number of partitions. 

As such, in the example as shown in Figure \ref{fig:figure1}(e), given that data streams are partitioned into three partitions, data stream $\mathcal{R}$ has a partition span equal to $\frac{8}{3}$ and will be partitioned into the following three partitions: \D{[1,3.67)}, \D{[3.67,6.33)} and \D{[6.33,9]}, and the same partitioning boundary calculations can also be applied to data stream $\mathcal{S}$ as indicated in line $2$-$3$ in Algorithm \ref{alg:alg_step2}. 

For all aforementioned algorithms, two data streams are partitioned separately based on their respective partitioning boundaries after they are obtained. As such there is no interference between the partitions of each stream, which implies the partitioning is accomplished in an isolated manner. Clearly, the isolated partitioning lacks the notion of collaborative partitioning information of the data streams, and thus damages the efficacy of the partition-level filtering. Take Figure \ref{fig:figure1}(e) as an example, for the coarse-grained partitioning, partition \D{[1,3.67)} of data stream $\mathcal{R}$ and partition \D{[2.67, 5.33)} of data stream $\mathcal{S}$ will be produced separately and the Cartesian products are performed between them under the ``$>$'' theta condition since valid theta-join results can be available between these two partitions. However, not all Cartesian products between these two partitions are necessary, say, the Cartesian products between \D{[1,2.67]} and \D{[3.67,5.33)} are completely useless as they are judged to be impossible to possess any valid theta-join results based on the ``$>$'' theta condition. Therefore, the coarse partitions produced by the coarse-grained isolated partitioning strategy will impair the efficacy of the partition-level filtering, which could result in more Cartesian products than necessary, and thus seriously hinder the efficiency of the theta-join operation. 

Given the drawback caused by this isolated coarse-grained partitioning strategy, we consider an amalgamated partitioning mechanism as shown in Algorithm \ref{alg:alg_step2}, which is useful to address this issue. To make the partitioning more fine-grained, after calculating the partitioning boundaries, we amalgamate the partitioning boundaries of data streams to form the amalgamated partitioning boundaries as presented in line $5$ in Algorithm \ref{alg:alg_step2}. By fusing the partitioning information of data streams, the aforementioned drawback can be circumvented. Therefore, the effectiveness of the partition-level filtering is improved, which would lead to the reduction of the number of Cartesian products to be conducted. 

In the example of Figure \ref{fig:figure1}(e), given the partitioning boundaries of data stream $\mathcal{R}$ are \D{[1,3.67)}, \D{[3.67,6.33)} and \D{[6.33,9]}, and the partitioning boundaries of data stream $\mathcal{S}$ are \D{[0,2.67)}, \D{[2.67,5.33)} and \D{[5.33,8]}, line 4-5 of Algorithm \ref{alg:alg_step2} will amalgamate these two partitioning boundaries together and the amalgamated partitioning boundaries produced in this case would be \D{[0,1)}, \D{[1,2.67)}, \D{[2.67,3.67)}, \D{[3.67,5.33)}, \D{[5.33,6.33)}, \D{[6.33,8)} and \D{[8,9]}. After applying the amalgamated partitioning scheme to both data streams, the Cartesian products in the above case between \D{[2.67,3.67)} from data stream $\mathcal{R}$ and \D{[2.67,3.67)} from data stream $\mathcal{S}$ will be conducted as usual, while the unnecessary Cartesian products between \D{[1,2.67]} from data stream $\mathcal{R}$ and \D{[3.67,5.33)} from data stream $\mathcal{S}$ will be avoided by the partition-level filtering thanks to the fine-grained amalgamated partitioning strategy. This significantly decreases the number of useless Cartesian products, and hence benefits the efficiency of the algorithm. 

\begin{algorithm*}[!ht]
  \begin{algorithmic}[1]
    \Require
      \StateX $\uptheta$ operator, $\uptheta$ $\in$ \{$<, \leq, >, \geq$\}, 
      \StateX window size $w$, 
      \StateX Amalgamated partitioned data stream $P_D, D \in \{\mathcal{R}, \mathcal{S}\}$
    \Ensure $\uptheta$-join results of two input data streams. 
    \State // \textbf{Step 3.1: Oversized Partition Re-partitioning}
    \State $AS_D$ $\leftarrow$ $\frac{w}{P_D.size()}$
    \For{each $P_D^i$ \textup{of $D$}}
      \If{$P_D^i$.size() $>$ $AS_D$}
        \State $sp^{i'}_D$ $\leftarrow$ $\lceil \frac{{P_D}^i_{max} - {P_D}^i_{min}}{rn^i_D} \rceil$
        \State Repartition $P_D^i$ as follows
        \StateX \hspace*{1.15em} [ ${P_D^i}_{min}$, ${P_D^i}_{min} + 1 \times sp^{i'}_D$ ), 
        \StateX \hspace*{1.15em} [ ${P_D^i}_{min} + 1 \times sp^{i'}_D$, ${P_D^i}_{min} + 2 \times sp^{i'}_D$ ), 
        \StateX \hspace*{8em} \vdots
        \StateX \hspace*{1.15em} [ ${P_D^i}_{min} + (rn^i_D - 1) \times sp^{i'}_D$, ${P_D^i}_{max}$ ]
      \EndIf
    \EndFor
    \State $P_D$ $\leftarrow$ $P_D$.filter($P_D^i$.size() != 0)
    \State // \textbf{Step 3.2: Partition-level Filtering}
    \For{\textup{each} $P_\mathcal{R}^i$ \textup{of} $P_\mathcal{R}$}
      \For{\textup{each} $P_\mathcal{S}^j$ \textup{of} $P_\mathcal{S}$}
        \If{$\uptheta$ \textup{is} ``$>$" \textup{and} ${P_\mathcal{R}^i}_{max} > {P_\mathcal{S}^j}_{min}$}
          \State Distribute Cartesian\_product($P_\mathcal{R}^i$, $P_\mathcal{S}^j$) to processors
        \ElsIf{$\uptheta$ \textup{is ``$\geq$"} \textup{and} ${P_\mathcal{R}^i}_{max} \geq {P_\mathcal{S}^j}_{min}$}
          \State Distribute Cartesian\_product($P_\mathcal{R}^i$, $P_\mathcal{S}^j$) to processors
        \ElsIf{$\uptheta$ \textup{is ``$<$"} \textup{and} ${P_\mathcal{R}^i}_{min} < {P_\mathcal{S}^j}_{max}$}
          \State Distribute Cartesian\_product($P_\mathcal{R}^i$, $P_\mathcal{S}^j$) to processors
        \ElsIf{$\uptheta$ \textup{is ``$\leq$"} \textup{and} ${P_\mathcal{R}^i}_{min} \leq {P_\mathcal{S}^j}_{max}$}
          \State Distribute Cartesian\_product($P_\mathcal{R}^i$, $P_\mathcal{S}^j$) to processors
        \EndIf
      \EndFor
    \EndFor
    \State // \textbf{Step 3.3: Cartesian Products and Theta-join Results}
    \For{all Cartesian products generated}
      \State Only keep those that satisfy the $\uptheta$ condition
    \EndFor
    \State \Return $\uptheta$-join results
  \end{algorithmic}
  \caption{\rw{The \emph{Prefap} algorithm - Auxiliary Procedures (Step 3)}}
  \label{alg:alg_step3}
\end{algorithm*}

\subsection{Auxiliary Procedures}
\label{sec:sec3.4_auxiliary_procedures}
To obtain the theta-join results, some auxiliary procedures as shown in Algorithm \ref{alg:alg_step3} that are adopted in the FastThetaJoin algorithm are exploited to follow up the proposed approaches in the \emph{Prefap} framework, which are detailed as follows: 

\textit{1) Oversized Partition Re-partitioning}: By following the common design scheme in \cite{fastthetajoin_original_hu2020fastthetajoin}, we design the framework that can re-partition any oversized partitions to balance the workload as much as possible in the distributed environment as shown in line $2$-$9$ in Algorithm \ref{alg:alg_step3}. Specifically, any partition with its size larger than the average partition size is regarded as an oversized partition. Once a partition is judged as an oversized partition, it will be re-partitioned into a number of sub-partitions defined as: 
\rw{\begin{equation}
  \label{equ:equation3_repartition_numer}
  rn^i_D = \ceil[\bigg]{\frac{P^i_D.size()}{AS_D}} = \ceil[\bigg]{\frac{P^i_D.size()}{\frac{w}{P_D.size()}}}, D \in \{\mathcal{R}, \mathcal{S}\}
\end{equation}}
\rw{and the pseudocode of this process is shown in line $3$-$8$ in Algorithm \ref{alg:alg_step3}. The oversized partition re-partitioning is completed by the Partitioning Unit of the \emph{Prefap} framework. Together with the fine-grained amalgamated partitioning, the load balancing of the framework can be improved. }

\textit{2) Partition-level Filtering}: \rw{Once both data streams are partitioned, the Partitioning Unit of the \emph{Prefap} framework will filter the partitions based on the theta condition as is presented in line $11$-$23$ in Algorithm \ref{alg:alg_step3}, so that the Cartesian products between the partitions that possess no valid theta-join results will not be performed. }

In the example illustrated in Figure \ref{fig:figure1}(e), even though partition \D{[1,2.67)} from data stream $\mathcal{R}$ and partition \D{[3.67,5.33)} from data stream $\mathcal{S}$ are both produced, the Cartesian products between these two partitions will not be performed under the ``$>$'' theta condition, because even the maximum value in the former partition is smaller than the minimum value in the latter partition, which is exactly line $13$-$14$ in Algorithm \ref{alg:alg_step3}. Hence, by applying the partition-level filtering based on the theta condition, unnecessary Cartesian products are eliminated, leading to a more efficient algorithm. 

\textit{3) Cartesian Products and Theta-join Results}: After completing the pre-filtering, amalgamated partitioning, oversized partition re-partitioning, and the partition-level filtering, the remaining Cartesian products are highly refined, and then the Cartesian products will be distributed to different processes to output the final theta-join results based on the theta condition as shown in line $24$-$28$ in Algorithm \ref{alg:alg_step3}. 


\section{Empirical Studies}
\label{sec:sec4_experiments_and_results_analysis}

To validate the effectiveness of the \emph{Prefap} algorithm, comprehensive evaluations are performed on both synthetic and real data streams from two-way to multi-way theta-join operations. We compare our algorithm against several theta-join algorithms, including the state-of-the-art algorithm FastThetaJoin (FTJ) \cite{fastthetajoin_original_hu2020fastthetajoin} and several well-known and widely used algorithms, such as Range-based Method (RBM) \cite{range_based_method_dewitt1992practical}, One-Bucket Theta (OBT) \cite{one_bucket_theta_okcan2011processing}, and Cross Filter Strategy (CFS) \cite{cross_filter_strategy_liu2017efficient}. 


\begin{table*}[!h]
  \caption{Theta-join performance evaluation metrics and their corresponding interpretation}
  \centering
  \resizebox{\textwidth}{!}{\begin{tabular}{c||c}
  \hline
  \bfseries Metric & \bfseries Interpretation\\
  \hline\hline
  Number of Cartesian products & The number of Cartesian products that the algorithm performs\\
  \hline
  Elapsed time & The total time elapsed, measured in milliseconds (ms)\\
  \hline
  Load balancing ratio (In) & The maximum input load among processes divided by the average input load among processes\\
  \hline
  Load balancing ratio (Out) & The maximum number of theta-join results among processes divided by the average number of theta-join results among processes\\
  \hline
  \end{tabular}}
  \label{tab:table1_performance_evaluation_metrics}
\end{table*}

\subsection{Experimental Setup and Data Streams}\label{sec:sec4.1_experimental_setup_and_datasets}

We follow the \emph{Prefap} workflow described in Section~\ref{sec:sec3.1_the_workflow_of_the_proposed_framework} to conduct the experiments where
the number of partitions $p$ is set to be $10$ and the window size $w$ to be $1000$. The evaluation metrics we use to evaluate the performance of the algorithms are listed in Table \ref{tab:table1_performance_evaluation_metrics} with their corresponding interpretations. To fully testify the effectiveness of the proposed algorithm, the theta-join on two-way data streams and multi-way data streams are performed to validate that the proposed algorithm scales well. The hypothesis testings are then conducted to demonstrate that the \emph{Prefap} algorithm achieves a statistically significant performance improvement compared with FTJ. The ablation studies are also conducted to show the efficacy of the algorithm and reveal the importance and necessity of each component introduced in the \emph{Prefap} framework. Finally, the number of partitions and the window size are adjusted to show the superior performance of the \emph{Prefap} algorithm in various settings. 

\begin{figure*}[!b]
  \begin{center}
    \includegraphics[height=3.6cm,keepaspectratio]{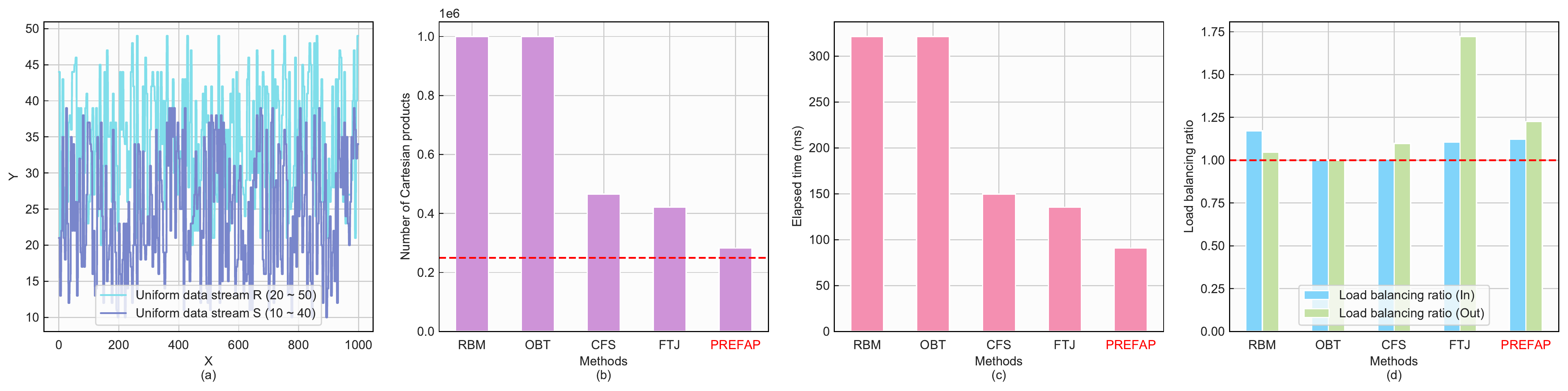}\\
    \caption{The performance comparisons between algorithms when performing theta-join on synthetic 2-way Uniform data streams. The theta condition is $\mathcal{R} \le \mathcal{S}$. The stream size of both data streams are 1000. The Uniform data stream $\mathcal{R}$ fluctuates in range $[20, 50]$, while the Uniform data stream $\mathcal{S}$ is in range $[10, 40]$. (a) depicts the 2-way data streams, (b), (c) and (d) present the number of Cartesian products, elapsed time and the in/out load balancing ratio, respectively. The red dotted line in (b) indicates the number of theta-join results, i.e., the minimum number of Cartesian products that need to be performed. The red dotted line in (d) marks 1.0, which indicates perfect load balancing. }
    \label{fig:figure3}
  \end{center}
\end{figure*}

\begin{figure*}[!b]
  \begin{center}
    \includegraphics[height=3.6cm,keepaspectratio]{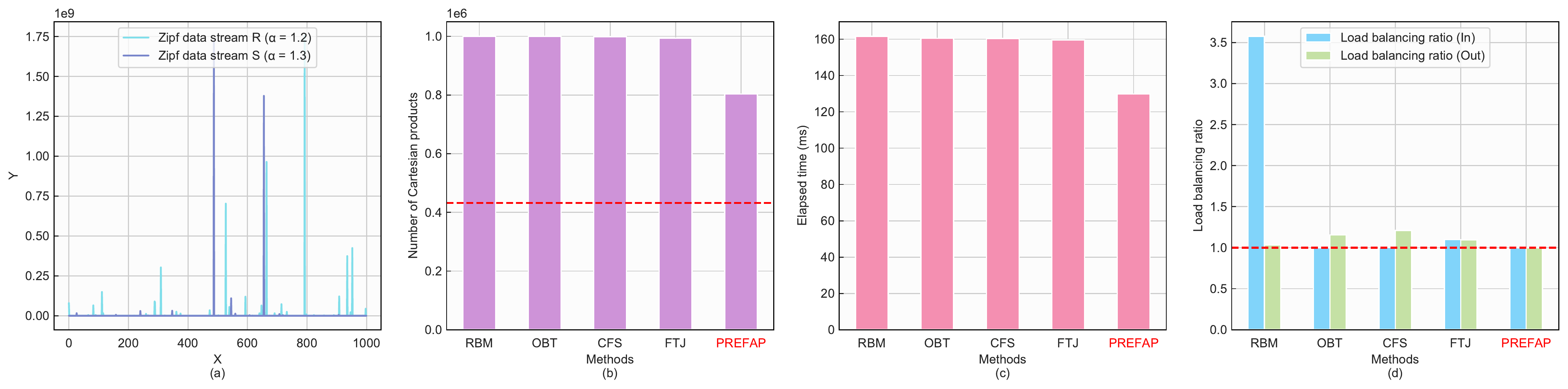}\\
    \caption{The performance comparisons between algorithms when performing theta-join on synthetic 2-way Zipf data streams. The theta condition is $\mathcal{R} \le \mathcal{S}$. The stream size of both data streams are 1000. The shape parameter $\alpha$ of Zipf data stream $\mathcal{R}$ and $\mathcal{S}$ are set to be $1.2$ and $1.3$, respectively. (a) depicts the 2-way data streams, (b), (c) and (d) present the number of Cartesian products, elapsed time and the in/out load balancing ratio, respectively. The red dotted line in (b) indicates the number of theta-join results, i.e., the minimum number of Cartesian products that need to be performed. The red dotted line in (d) marks 1.0, which indicates perfect load balancing. }
    \label{fig:figure4}
    \vspace{-5mm}
  \end{center}
\end{figure*}

We use both synthetic and real data streams to testify the effectiveness of the proposed algorithm. For the synthetic datasets, randomly generated \emph{Uniform} and \emph{Normal} data streams are exploited to represent some typical non-skewed data streams, while the randomly generated \emph{Zipf} data stream acts as a representative of a typical skewed data stream. As for the real datasets, by following the method in \cite{cross_filter_strategy_liu2017efficient}, we use the Clouds dataset provided by the U.S. Department of Energy in our experiments where the real-time wind speed measured in metre per second ($m/s$) of different months in $2000$ are leveraged to form different data streams. Additionally, the stock market price data streams provided by Yahoo! Finance \cite{yahoo_finance2021} are also utilised, in which the real-time high price of stocks of different companies between $2010$ and $2020$ are used to form different data streams. The detailed parameter settings of different synthetic data streams will be given when they are used in the following subsections. 

In terms of the experimental infrastructure, a server equipped with Intel Core i7 7600U CPU and 32GB of memory is utilised. The CPU has two cores with 4 processors, so that the Cartesian products involved in the \emph{Prefap} framework can be computed in a distributed manner. 

\begin{figure*}[!t]
  \begin{center}
    \includegraphics[height=3.6cm,keepaspectratio]{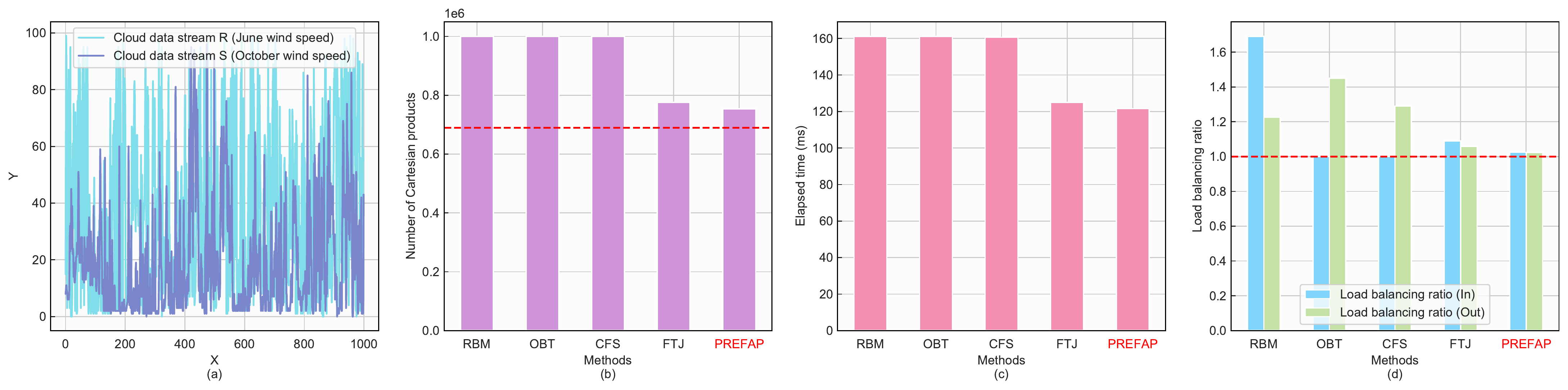}\\
    \caption{The performance comparisons between algorithms when performing theta-join on real 2-way Clouds data streams. The theta condition is $\mathcal{R} \ge \mathcal{S}$. The stream size of both data streams are 1000. The Clouds data stream $\mathcal{R}$ and $\mathcal{S}$ are the real-time wind speed captured in every 5 seconds in June 2000 and October 2000, respectively. Note that these two months are randomly selected as representatives. (a) depicts the 2-way data streams, (b), (c) and (d) present the number of Cartesian products, elapsed time and the in/out load balancing ratio, respectively. The red dotted line in (b) indicates the number of theta-join results, i.e., the minimum number of Cartesian products that need to be performed. The red dotted line in (d) marks 1.0, which indicates perfect load balancing. }
    \label{fig:figure5}
  \end{center}
\end{figure*}

\begin{figure*}[!t]
  \begin{center}
    \includegraphics[height=3.6cm,keepaspectratio]{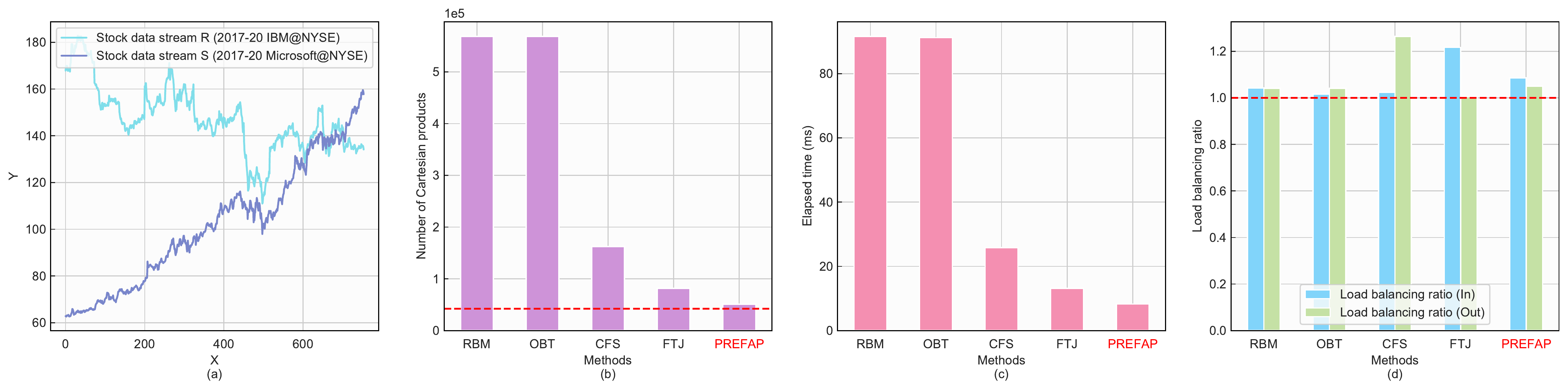}\\
    \caption{The performance comparisons between algorithms when performing theta-join on real 2-way Stock data streams. The theta condition is $\mathcal{R} \le \mathcal{S}$. The stream size of both data streams are 755. The Stock data stream $\mathcal{R}$ and $\mathcal{S}$ are the New York Stock Exchange (NYSE) stock high price recorded daily between $3$\textsuperscript{rd} Jan 2017 and $2$\textsuperscript{nd} Jan 2020 of company IBM and Microsoft, respectively. Note that there are only 755 stock exchange open days in this period, and these two companies are randomly selected as representatives. (a) depicts the 2-way data streams, (b), (c) and (d) present the number of Cartesian products, elapsed time and the in/out load balancing ratio, respectively. The red dotted line in (b) indicates the number of theta-join results, i.e., the minimum number of Cartesian products that need to be performed. The red dotted line in (d) marks 1.0, which indicates perfect load balancing. }
    \label{fig:figure6}
  \end{center}
  \vspace{-5mm}
\end{figure*}

\subsection{2-way Data Stream Theta-join}
\label{sec:sec4.2_results_on_2way_data_stream_theta_join}

The performances of theta-join algorithms for two-way uniform data streams, zipf data streams, clouds data streams, and stock price data streams are illustrated in Figure \ref{fig:figure3} to \ref{fig:figure6}, respectively. Note that the detailed data stream and distribution configurations (including stream size, distribution parameters, join attribute used, $\theta$ condition) have been given in the corresponding image captions. 

The results show that, in all cases, even for the highly skewed data streams such as the zipf data streams, the \emph{Prefap} algorithm performs better than the state-of-the-art method FTJ in terms of efficiency, and significantly outperforms all other algorithms. More specifically, the \emph{Prefap} algorithm achieves $34.7\%$, $19.4\%$, $3.0\%$ and $37.7\%$ reductions on the number of performed Cartesian products compared with FTJ with respect to $4$ different kinds of data streams, respectively. The significant reduction of the number of Cartesian products yields better efficiency, as indicated by the shortest elapsed time performance in all cases. 

The red dotted lines in Figure \ref{fig:figure3} (b) to \ref{fig:figure6}(b) indicate the number of theta-join results, i.e., no matter how the algorithm is optimised, it must perform at least this number of Cartesian products to yield the complete result. As shown in the results, the \emph{Prefap} algorithm significantly minimises the gap between the number of Cartesian products it performs and the optimal case. Compared with FTJ, it clearly indicates an effective performance boost and thus demonstrates that the cooperation between the pre-filtering strategy and the amalgamated partitioning mechanism contributes positively towards reducing the redundancy in Cartesian products, which in turn significantly boosts the performance. 

In terms of the load balancing as illustrated in Figure \ref{fig:figure3} to \ref{fig:figure6}, compared with the FTJ, the \emph{Prefap} algorithm performs better for both input and output load balancing ratios, indicated by the red dotted line in sub-figure (d). Hence, it justifies that with the collaboration between the fine-grained amalgamated partitioning scheme and the oversized partition re-partitioning mechanism, the \emph{Prefap} algorithm performs well in the distributed environment by distributing workloads in a relatively even way. 

\begin{figure*}[!b]
  \begin{center}
    \includegraphics[width=\textwidth,keepaspectratio]{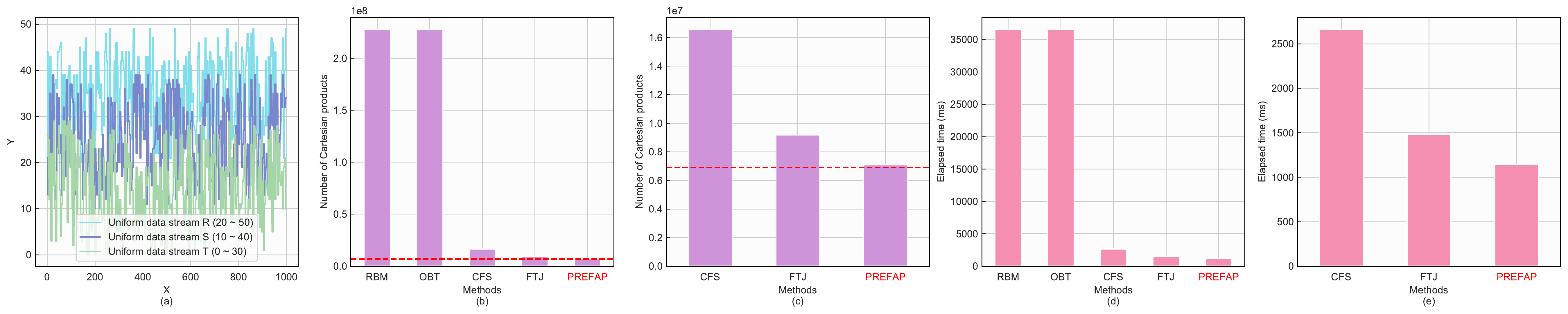}\\
    \caption{The performance comparisons between algorithms when performing theta-join on synthetic 3-way Uniform data streams. The theta condition is $\mathcal{R} < \mathcal{S} \le \mathcal{T}$. The stream size of all three data streams are 1000. The Uniform data stream $\mathcal{R}$, $\mathcal{S}$ and $\mathcal{T}$ fluctuate in range $[20, 50]$, $[10, 40]$ and $[0, 30]$, respectively. (a) depicts the 3-way data streams, (b) and (d) present the number of Cartesian products and the elapsed time, respectively. To clearly show the performance gain, (c) and (e) are the zoom-in version of (b) and (d), respectively. Both RBM and OBT are omitted due to their worst performances. The red dotted lines in (b) and (c) indicate the number of theta-join results, i.e., the minimum number of Cartesian products that need to be performed. }
    \label{fig:figure7}
  \end{center}
\end{figure*}

\begin{figure*}[!b]
  \begin{center}
    \includegraphics[width=\textwidth,keepaspectratio]{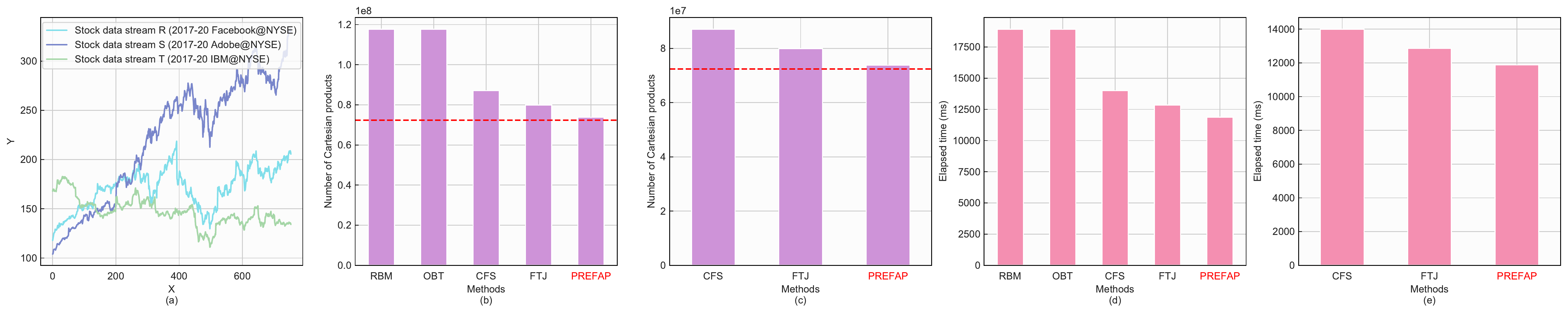}\\
    \caption{The performance comparisons between algorithms when performing theta-join on real 3-way Stock data streams. The theta condition is $\mathcal{R} \ge \mathcal{S} < \mathcal{T}$. The stream size of all three data streams are 755. The Stock data stream $\mathcal{R}$, $\mathcal{S}$ and $\mathcal{T}$ are the New York Stock Exchange (NYSE) stock high price recorded daily between $3$\textsuperscript{rd} Jan 2017 and $2$\textsuperscript{nd} Jan 2020 of company Facebook, Adobe and IBM, respectively. Note that there are only 755 stock exchange open days in this period, and these three companies are randomly selected as representatives. (a) depicts the 3-way data streams, (b) and (d) present the number of Cartesian products and the elapsed time, respectively. To clearly show the performance gain, (c) and (e) are the zoom-in version of (b) and (d), respectively. Both RBM and OBT are omitted due to their worst performances. The red dotted lines in (b) and (c) indicate the number of theta-join results, i.e., the minimum number of Cartesian products that need to be performed. }
    \label{fig:figure8}
  \end{center}
\end{figure*}

\subsection{Multi-way Data Stream Theta-join}\label{sec:sec4.3_results_on_multiway_data_stream_theta_join}

To further verify that the \emph{Prefap} algorithm scales well when processing multi-way data streams, comprehensive experiments are conducted on both synthetic and real multi-way data streams. As shown in Figure \ref{fig:figure7}(a), the uniform multi-way data streams represent a typical example of the non-skewed data streams. While as illustrated in Figure \ref{fig:figure8}(a), the New York Stock Exchange (NYSE) stock prices of three randomly selected companies between $2017$ and $2020$ are skewed and are therefore capable of fully testifying the effectiveness of the proposed algorithm. Note that the detailed data stream configurations have been given in the corresponding image captions.

In Figure \ref{fig:figure7} to \ref{fig:figure8}, sub-figure (b) and (d) present the evaluation results of the number of Cartesian products and the elapsed time, respectively. As the gaps between the \emph{Prefap} algorithms and other algorithms are so significant, the sub-figure (b) and (d) have been zoomed in as shown in sub-figure (c) and (e), respectively. As we can see, under both synthetic and real multi-way data streams, the \emph{Prefap} algorithm outperforms the FTJ algorithm by a large margin, achieving a $24.1\%$ and a $7.7\%$ decrease in the number of Cartesian products in two cases, respectively. Hence, it is also natural to observe that the elapsed time in both cases are reduced accordingly, hence the algorithm becomes more efficient. Furthermore, the \emph{Prefap} algorithm attains a near-optimal performance. Given the number of theta-join results, i.e. the minimum number of Cartesian products required to be performed that is indicated by the red dotted line in sub-figure (b) and (c) in Figure \ref{fig:figure7} to \ref{fig:figure8}, the \emph{Prefap} algorithm only performs $0.026\%$ and $1.7\%$ more Cartesian products compared with the optimal scenario, indicating excellent theta-join performance. 

Therefore, the superior performance of the \emph{Prefap} algorithm in various cases strongly testifies that the collaboration of the pre-filtering strategy and the amalgamated partitioning mechanism is effective in improving the theta-join performance. 

\begin{figure*}[!b]
  \begin{center}
    \includegraphics[height=3.6cm,keepaspectratio]{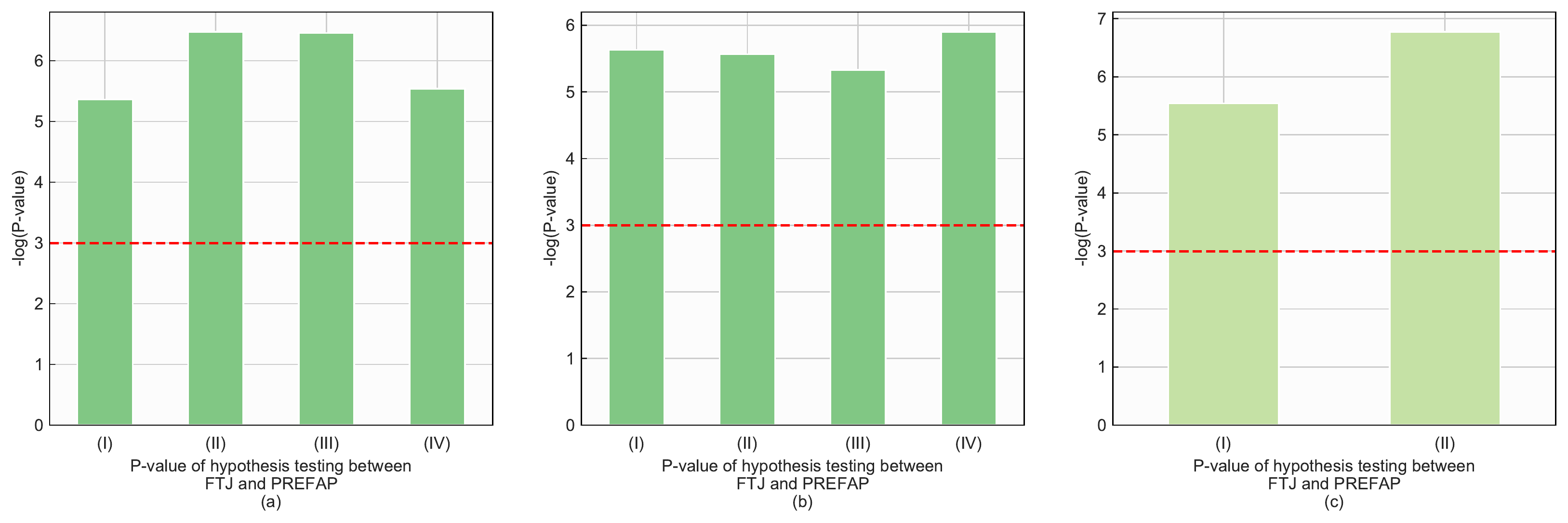}\\
    \caption{The one-sided significance T-tests with $0.05$ as the significance level are conducted under three tasks to testify the performance gains of the \emph{Prefap} algorithm compared with the best compared method FastThetaJoin (FTJ). The y-axis denotes the $-log(p\_value)$, and the red dotted lines in all three subplots mark the significance threshold, i.e., $-log(0.05)$. (a), (b) and (c) are for theta-join on synthetic 2-way Zipf data streams (stream size = $1000$, $\alpha = 1.2, 1.3$, $\theta: \mathcal{R} \le \mathcal{S}$), real 2-way Clouds data streams (stream size = $1000$, $\theta: \mathcal{R} \ge \mathcal{S}$, data stream $\mathcal{R}$ and $\mathcal{S}$ are the real-time wind speed captured in every 5 seconds in June 2000 and October 2000, respectively.), and real 3-way Stock data streams (stream size = $755$, $\theta: \mathcal{R} \ge \mathcal{S} < \mathcal{T}$, data stream $\mathcal{R}$, $\mathcal{S}$ and $\mathcal{T}$ are the New York Stock Exchange (NYSE) stock high price recorded daily between 3rd Jan 2017 and 2nd Jan 2020 of company Facebook, Adobe and IBM, respectively. ). (I), (II), (III) and (IV) indicates the significance of the number of Cartesian products, elapsed time, load balancing ratio (In) and load balancing ratio (Out). If the bar is higher than the red dotted line, it indicates that the alternate hypothesis is accepted, i.e., the \emph{Prefap} algorithm outperforms the FastThetaJoin algorithm on that metric. The higher the bar is, the more significant the performance improvement produced by the \emph{Prefap} algorithm is. }
    \label{fig:figure9}
  \end{center}
\end{figure*}

\subsection{Significance Tests}\label{sec:sec4.4_hypothsis_testing}

To further validate that the improvement achieved by \emph{Prefap} over the state-of-the-art FTJ algorithm is statistically significant, three representative data stream settings, i.e., theta-join between 2-way zipf data streams, 2-way cloud data streams, and multi-way stock price data streams are performed $30$ times and the \emph{T-tests} are employed to verify the significance of the performance boost using $0.05$ as the significance threshold. Note that the detailed data stream configuration has been given in the image caption. 

As shown in Figure \ref{fig:figure9}, $4$ bars I, II, III, IV in sub-figure (a) and (b) indicate the $-\log(p\_value)$ of the one-sided T-test on 4 different evaluation metrics as presented in Table \ref{tab:table1_performance_evaluation_metrics} for two 2-way theta-join settings. The $-\log(p\_value)$ results of the multi-way theta-join are presented in sub-figure(c), in which the bars correspond to the $-\log(p\_value)$ results of the differences between \emph{Prefap} and FTJ on the number of Cartesian products and the elapsed time, respectively. In Figure \ref{fig:figure9}, the red dotted lines indicate $-\log(0.05)$, which is the significance threshold of the one-sided T-test. If the bar is higher than the red dotted line, it indicates the acceptance of the alternate hypothesis, which is Metric$_\mathit{Prefap} <$ Metric$_\mathit{FTJ}$. And the higher the bar is, the more significant the performance improvement produced by the \emph{Prefap} algorithm is.

From Figure \ref{fig:figure9}, we can clearly observe that all bars in all cases are significantly higher than the red dotted line, which indicates that in all data stream settings, the \emph{Prefap} algorithm achieves a lower evaluation metric value than the FTJ algorithm, i.e., less number of Cartesian products are performed, less amount of elapsed time is required, and the load balancing ratio of both input and output is closer to $1$. The heights of the bars are much higher than the red dotted line, which indicate superior performances with high statistical confidence. Hence, statistically, the T-test results verify the superiority of the \emph{Prefap} algorithm, and the performance enhancement is statistically significant. The statistically testifiable superior performance also demonstrates that the \emph{Prefap} framework can collaboratively contribute towards better theta-join efficiency and excellent load balancing in the distributed environment, the benefit brings by the \emph{Prefap} framework is statistically significant. 

\subsection{Ablation Study}\label{sec:sec4.5_ablation_study}

The ablation study is conducted by evaluating several variants of the \emph{Prefap} framework: (1) \emph{Prefap} with the pre-filtering being removed; (2) \emph{Prefap} with the amalgamated partitioning being turned off; and (3) \emph{Prefap} with both the pre-filtering and the amalgamated partitioning being ablated. 

Two-way and multi-way stock price data streams of randomly selected companies are used as the representative tasks on which the theta-join is performed by using both the full \emph{Prefap} algorithm and its ablated variants. The results are presented in Table \ref{tab:table2} and \ref{tab:table3} for two-way and multi-way theta-join tasks, respectively. Note that the detailed data stream configuration is also provided in table captions. We can observe that the full \emph{Prefap} algorithm outperforms all its variants by a large margin, which indicates that any one of these components in the \emph{Prefap} framework plays an indispensable role and brings benefits to reduce the number of Cartesian products being performed and enhance the overall efficiency. Among all these components, the amalgamated partitioning scheme brings the highest amount of Cartesian product reductions in two tasks, which is $16.8\%$ and $6.3\%$, respectively. Correspondingly, the elapsed time drops by $17.5\%$ and $6.2\%$, respectively. This further validates the importance of amalgamating the partitioning scheme and the benefits of avoiding coarse-grained isolated partitioning. As indicated in Table \ref{tab:table2} and \ref{tab:table3}, although the pre-filtering strategy attains relatively less amount of performance gain, which is $1.0\%$ and $0.2\%$ in the two-way and multi-way task, respectively, however, the collaboration of both the pre-filtering and the amalgamated partitioning yields a tremendous decrease of Cartesian products of $29.1\%$ and $12.9\%$, respectively, for these two tasks. Therefore, by collaborating the pre-filtering strategy with the amalgamated partitioning mechanism in the \emph{Prefap} framework, it is highly effective in improving the efficiency of the theta-join operation. As such, the promising results successfully verify the effectiveness of the \emph{Prefap} algorithm. 

\begin{table*}[!h]
  \caption{Ablation study of the \emph{Prefap} algorithm performed on real 2-way Stock data streams ($\theta: \mathcal{R} \le \mathcal{S}$, $\mathcal{R}$: 2017-20 Johnson\&Johnson@NYSE, $\mathcal{S}$: 2017-20 Microsoft@NYSE, stream size = $755$, joined attribute: stock high price, symbol \textcolor{orange_}{\xmark} \ represents ``remove'')}
  \centering
  \resizebox{0.8\textwidth}{!}{\begin{tabular}{c||c|c}
  \hline
  \bfseries Setting & \bfseries Number of Cartesian products & \bfseries Elapsed time (ms)\\
  \hline\hline
  \textbf{Full version} & \textbf{104,138} & \textbf{33.68}\\
  \hline
  \textcolor{orange_}{\xmark} \ Pre-filtering & 105,132 \ \textcolor{blue}{($1.0\% \uparrow$)} & 33.89 \ \textcolor{blue}{($0.6\% \uparrow$)}\\
  \hline
  \textcolor{orange_}{\xmark} \ Amalgamated partitioning & 121,647 \ \textcolor{blue}{($16.8\% \uparrow$)} & 39.58 \ \textcolor{blue}{($17.5\% \uparrow$)}\\
  \hline
  \textcolor{orange_}{\xmark} \ Pre-filtering \& \textcolor{orange_}{\xmark} \ Amalgamated partitioning & 134,449 \ \textcolor{purple_}{($29.1\% \uparrow$)} & 43.03 \ \textcolor{purple_}{($27.8\% \uparrow$)}\\
  \hline\hline
  $\#$ of $\theta$-join results (best performance possible) & 92,240 & $--$\\
  \hline
  \end{tabular}}
  \label{tab:table2}
\end{table*}

\begin{table*}[!h]
  \caption{Ablation study of the \emph{Prefap} algorithm performed on real 3-way Stock data streams ($\theta: \mathcal{R} \ge \mathcal{S} < \mathcal{T}$, $\mathcal{R}$: 2017-20 Facebook@NYSE, $\mathcal{S}$: 2017-20 Adobe@NYSE, $\mathcal{T}$: 2017-20 IBM@NYSE, stream size = $755$, joined attribute: stock high price, symbol \textcolor{orange_}{\xmark} \ represents ``remove'')}
  \centering
  \resizebox{0.8\textwidth}{!}{\begin{tabular}{c||c|c}
  \hline
  \bfseries Setting & \bfseries Number of Cartesian products & \bfseries Elapsed time (ms)\\
  \hline\hline
  \textbf{Full version} & \textbf{74,923,241} & \textbf{12087.37}\\
  \hline
  \textcolor{orange_}{\xmark} \ Pre-filtering & 75,093,479 \ \textcolor{blue}{($0.2\% \uparrow$)} & 12110.81 \ \textcolor{blue}{($0.2\% \uparrow$)}\\
  \hline
  \textcolor{orange_}{\xmark} \ Amalgamated partitioning & 79,619,026 \ \textcolor{blue}{($6.3\% \uparrow$)} & 12838.10 \ \textcolor{blue}{($6.2\% \uparrow$)}\\
  \hline
  \textcolor{orange_}{\xmark} \ Pre-filtering \& \textcolor{orange_}{\xmark} \ Amalgamated partitioning & 84,558,245 \ \textcolor{purple_}{($12.9\% \uparrow$)} & 13628.14 \ \textcolor{purple_}{($12.7\% \uparrow$)}\\
  \hline\hline
  $\#$ of $\theta$-join results (best performance possible) & 72,510,414 & $--$\\
  \hline
  \end{tabular}}
  \label{tab:table3}
\end{table*}

\begin{figure*}[!htb]
  \begin{center}
    \includegraphics[height=3.9cm,keepaspectratio]{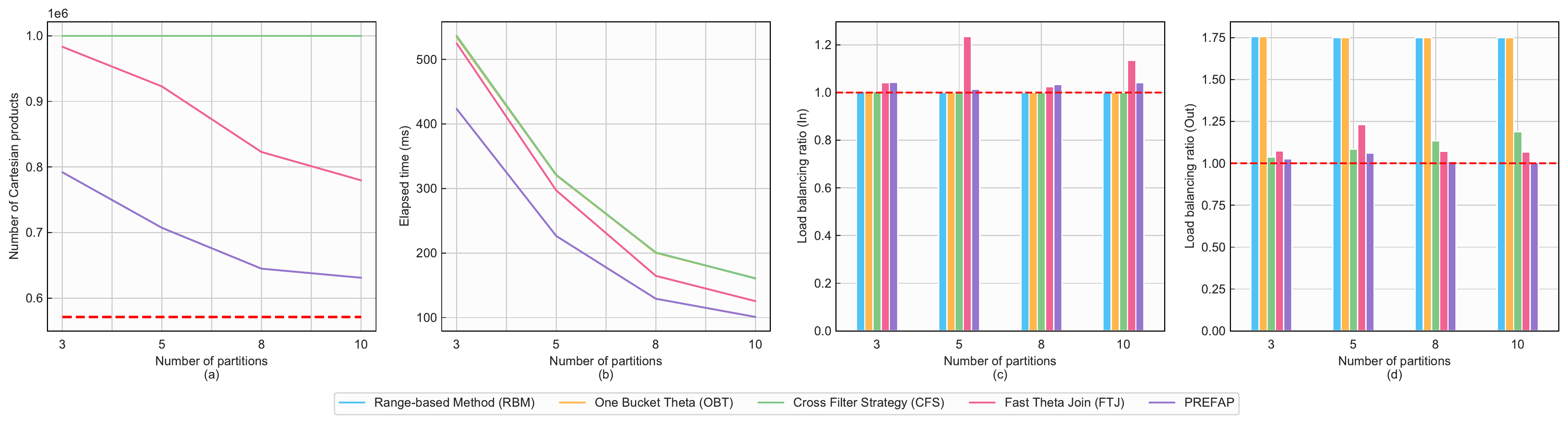}\\
    \caption{The performance comparisons between algorithms when performing theta-join on synthetic 2-way Normal data streams ($\mu = 1.2, 1, \sigma = 1, 1, \theta: \mathcal{R} > \mathcal{S}$, stream size = $1000$) under different number of partitions ($3$, $5$, $8$ and $10$). (a), (b), (c) and (d) present the number of Cartesian products, elapsed time, load balancing ratio (In) and load balancing ratio (Out), respectively. The red dotted line in (a) indicates the number of theta-join results, i.e., the minimum number of Cartesian products that need to be performed. The red dotted lines in (c) and (d) mark 1.0, which indicates perfect load balancing. }
    \label{fig:figure10}
  \end{center}
\end{figure*}

\begin{figure*}[!htb]
  \begin{center}
    \includegraphics[height=3.9cm,keepaspectratio]{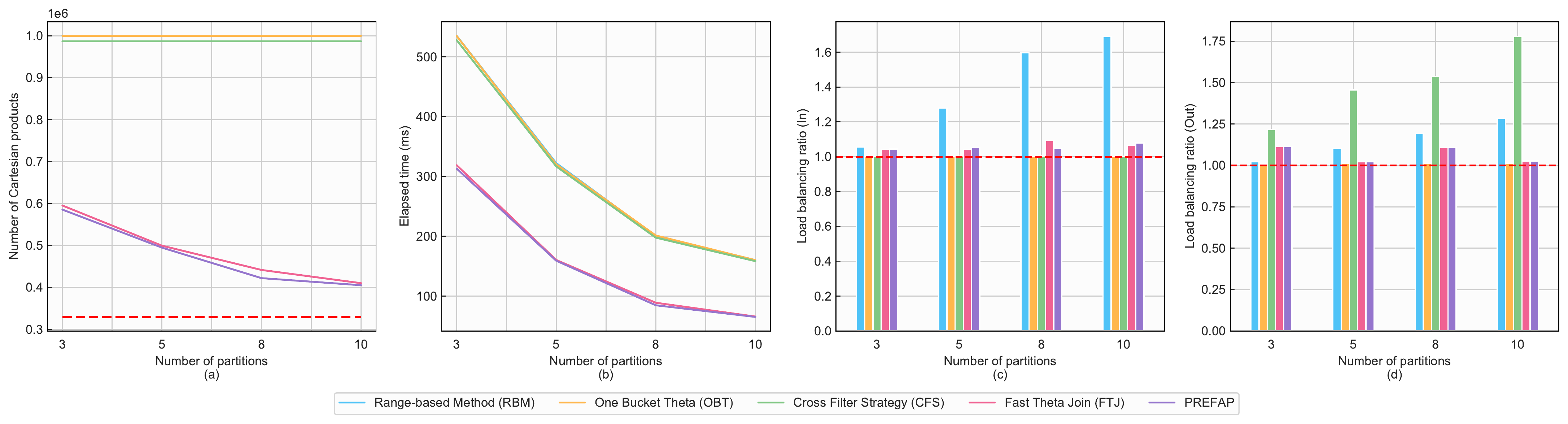}\\
    \caption{The performance comparisons between algorithms when performing theta-join on real 2-way Clouds data streams ($\theta: \mathcal{R} \le \mathcal{S}$, stream size = $1000$. The Clouds data stream $\mathcal{R}$ and $\mathcal{S}$ are the real-time wind speed captured in every 5 seconds in June 2000 and October 2000, respectively. ) under different number of partitions ($3$, $5$, $8$ and $10$). (a), (b), (c) and (d) present the number of Cartesian products, elapsed time, load balancing ratio (In) and load balancing ratio (Out), respectively. The red dotted line in (a) indicates the number of theta-join results, i.e., the minimum number of Cartesian products that need to be performed. The red dotted lines in (c) and (d) mark 1.0, which indicates perfect load balancing.}
    \label{fig:figure11}
  \end{center}
\end{figure*}

\vspace{-1.0cm}

\begin{figure*}[t]
  \begin{center}
    \includegraphics[height=3.9cm,keepaspectratio]{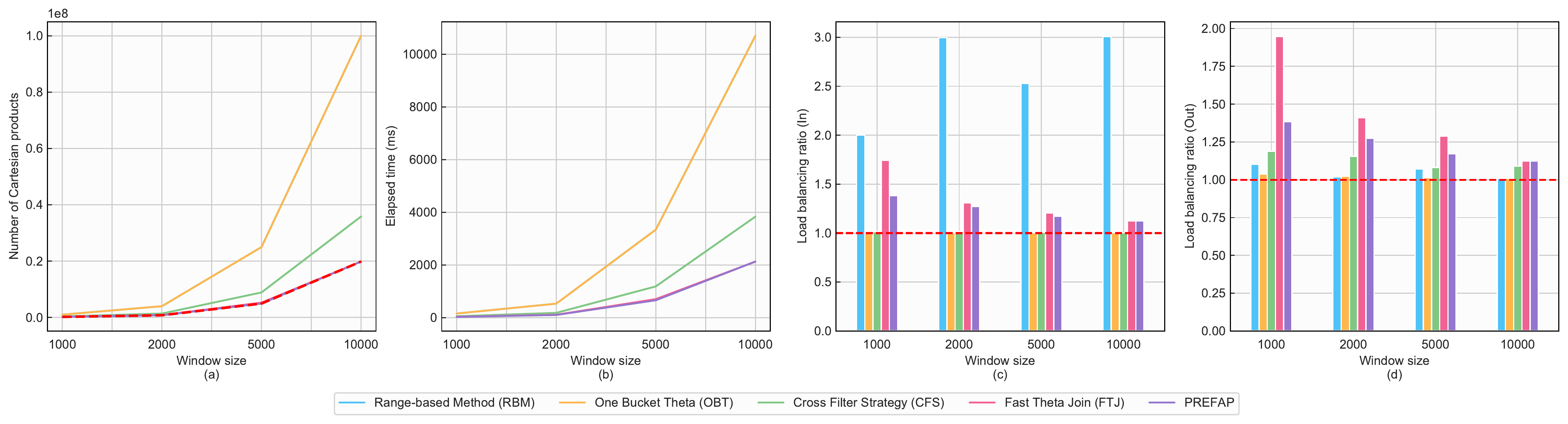}\\
    \caption{The performance comparisons between algorithms when performing theta-join on synthetic 2-way Uniform data streams ($range \in [0, 15], [5, 20], \theta: \mathcal{R} > \mathcal{S}$) in different window sizes ($1000$, $2000$, $5000$ and $10000$). (a), (b), (c) and (d) present the number of Cartesian products, elapsed time, load balancing ratio (In) and load balancing ratio (Out), respectively. The red dotted line in (a) indicates the number of theta-join results, i.e., the minimum number of Cartesian products that need to be performed. The red dotted lines in (c) and (d) mark 1.0, which indicates perfect load balancing. }
    \label{fig:figure12}
  \end{center}
  \vspace{-5mm}
\end{figure*}

\begin{figure*}[!htb]
  \begin{center}
    \includegraphics[height=3.9cm,keepaspectratio]{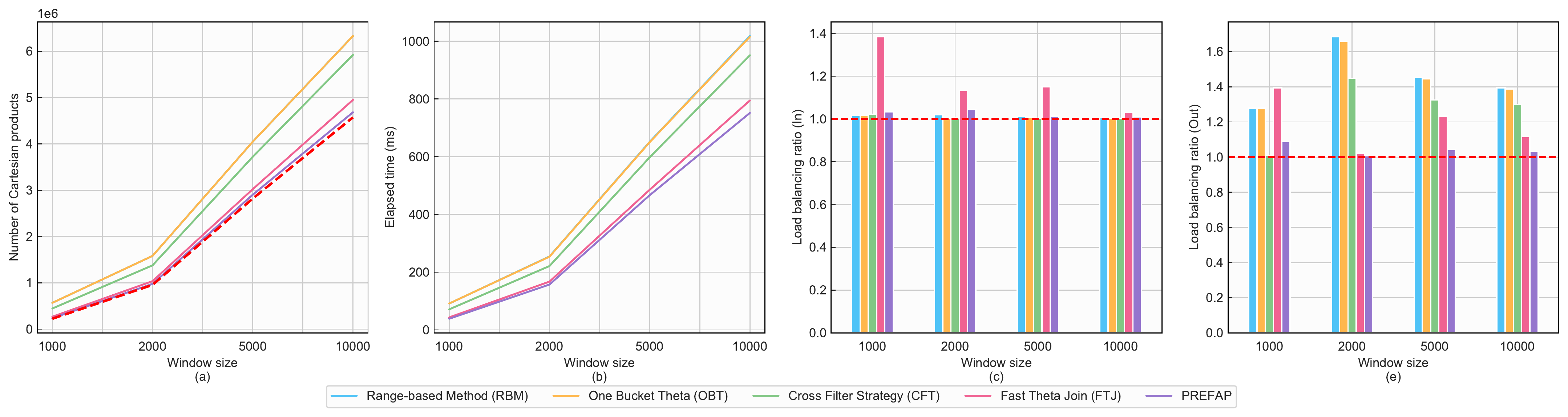}\\
    \caption{The performance comparisons between algorithms when performing theta-join on real 2-way Stock data streams ($\theta: \mathcal{R} > \mathcal{S}$, $\mathcal{R}$: 2010-20 FedExpress@NYSE, $\mathcal{S}$: 2010-20 Adobe@NYSE) in different window sizes ($1000$, $2000$, $5000$ and $10000$). (a), (b), (c) and (d) present the number of Cartesian products, elapsed time, load balancing ratio (In) and load balancing ratio (Out), respectively. The red dotted line in (a) indicates the number of theta-join results, i.e., the minimum number of Cartesian products that need to be performed. The red dotted lines in (c) and (d) mark $1.0$, which indicates perfect load balancing. }
    \label{fig:figure13}
  \end{center}
\end{figure*}

\subsection{Algorithm Efficiency}
\label{sec:sec4.6_partitions}

\textit{Number of Partitions}: To verify the effectiveness of the \emph{Prefap} algorithm when working with different number of partitions, the performance of the \emph{Prefap} algorithm with different number of partitions when processing two randomly selected tasks are presented in Figure \ref{fig:figure10} and \ref{fig:figure11}. The detailed data stream configuration has been given in the corresponding image caption. For both tasks, the number of Cartesian products decreases with the increase of the number of partitions, thanks to the more fine-grained partitioning effect brought by a larger number of partitions, as it benefits the filtering strategy and makes it more effective. In all partition settings, the \emph{Prefap} algorithm outperforms the FTJ algorithm by significantly reducing the number of Cartesian products being performed, and the elapsed time is reduced accordingly. This further demonstrates the superiority of the \emph{Prefap} algorithm with different number of partitions. 

Also, when the number of partitions is raised, the input and output load balancing ratio are improved marginally and are relatively close to $1$, which indicates a relatively even workload distribution in the distributed environment achieved by the \emph{Prefap} algorithm. 

\label{sec:sec4.7_window_sizes}
\textit{Window Sizes}: The performance of the \emph{Prefap} is also evaluated when the window size is varied. The performance on two randomly selected tasks, i.e., 2-way uniform data streams and 2-way stock price data streams, are shown in Figure \ref{fig:figure12} and \ref{fig:figure13}, respectively. The detailed data stream configuration has been given in the corresponding image caption. 

According to the experimental results, the number of Cartesian products being conducted and the elapsed time grow relatively proportionally with the increase of the window size, which demonstrates that the \emph{Prefap} algorithm scales well in terms of window size. In all window size settings, the \emph{Prefap} algorithm attains the lowest number of Cartesian products compared with its counterparts, and is relatively close to the optimal case marked by the red dotted line in sub-figure (a). The excellent performance of Cartesian product reduction benefits the efficiency of the algorithm, as indicated by the lowest elapsed time achieved by the \emph{Prefap} algorithm. 

Meanwhile, from sub-figure (c) and (d) of Figure \ref{fig:figure12} and \ref{fig:figure13}, the \emph{Prefap} algorithm attains a superior performance, compared with the FTJ algorithm, on both the input and the output load balancing ratio in all window size settings. Hence, it further validates the excellent scalability of the \emph{Prefap} algorithm in terms of window size. 

\section{Conclusion}
\label{sec:sec5_conclusion}

In this paper, we propose the \emph{Prefap} algorithm to enhance the performance of the theta-join operation. Compared with the FastThetaJoin algorithm, the pre-filtering strategy is applied to filter data elements in data streams that are deemed not possible to produce any valid theta-join results. The pre-filtering not only reduces the amount of data and hence lessens the workload, but also makes the partitioning more fine-grained to benefit further filtering. Then, during partitioning, the amalgamated partitioning scheme is employed to amalgamate the partitioning of two data streams, so that the performance degradation of the partition-level filtering caused by the coarse-grained isolated partitioning is avoided. By collaborating these mechanisms with the oversized partition re-partitioning strategy, as well as the partition-level filtering mechanism based on the theta condition, it forms our proposed \emph{Prefap} framework and it becomes more efficient when performing the theta-join operation. Comprehensive experiments and analyses are conducted to demonstrate the superiority of the performance against the recently published FastThetaJoin and several well-known theta-join algorithms. 

\section*{Acknowledgment}

This work is supported in part by Key-Area Research and Development Program of Guangdong Province (2020B010164002) and Shenzhen Basic Research Program (No. JCYJ20170818153016513 and JCYJ20200109115418592). 

\section*{Data Availability Statement}

The data that support the findings of this study are available from the corresponding author upon reasonable request. 

\bibliography{Prefap}

\end{document}